\documentclass[12pt]{article}
\usepackage{amsmath,comment}
\usepackage{graphicx}
\usepackage{enumerate}
\usepackage{natbib}
\usepackage{url} % not crucial - just used below for the URL 

\usepackage[utf8]{inputenc}
\usepackage{pgf, tikz}
\usepackage{adjustbox}
\usepackage{xcolor,color}
\usepackage{algorithm}
\usepackage{algpseudocode}
\usepackage{pgfpages}
\usepackage{subfigure}
\usepackage{ulem}
\usepackage{blindtext}
\graphicspath{{plots/}}
\usepackage{subfiles} % Best loaded last in the preamble
\usepackage{float}
\usepackage[color]{changebar}
\cbcolor{blue}
\RequirePackage{amsthm,amsmath,amsfonts,amssymb}
\usetikzlibrary{arrows, automata}
\usepackage{amsmath, amsthm, amssymb, bm, bbm}
\usepackage{amsfonts}
\usepackage{enumitem}
\usepackage{mathrsfs}

\usepackage[colorlinks = true,
            linkcolor = blue,
            urlcolor  = blue,
            citecolor = blue,
            anchorcolor = blue]{hyperref}
\usepackage{cleveref}

\newcommand{\E}{\mathbb{E}}

\newcommand{\V}{\mathcal{V}}
\newcommand{\OO}{\mathcal{O}}

\newcommand{\vect}[1]{\mathbf{#1}}
\newcommand{\norm}[1]{||{#1}||}
\newcommand{\R}{\mathbb{R} }

\newcommand{\al}[1]{\begin{align*}#1\end{align*}}

\newcommand{\Cov}{\mathrm{Cov}}
\newcommand{\Var}{\mathrm{Var}}

\newcommand{\indep}{\perp \!\!\! \perp}

\DeclareMathOperator*{\argmin}{arg\,min}

\newtheorem{theorem}{Theorem}[section]
\newtheorem{lemma}[theorem]{Lemma}
\newtheorem{assumption}[theorem]{Assumption}
\newtheorem{definition}[theorem]{Definition}

\newtheorem{example}{Example}[section]

\let\hat\widehat
\let\tilde\widetilde
\let\bar\overline
\let\gets\leftarrow

\newcommand{\blind}{1}

\begin{document}

\def\spacingset#1{\renewcommand{\baselinestretch}%
{#1}\small\normalsize} \spacingset{1}

%%%%%%%%%%%%%%%%%%%%%%%%%%%%%%%%%%%%%%%%%%%%%%%%%%%%%%%%%%%%%%%%%%%%%%%%%%%%%%

\if1\blind
{
  \title{\bf Model-Based Inference and Experimental Design for Interference Using Partial Network Data}
  \author{Steven Wilkins Reeves \hspace{.2cm}\\
  Department of Statistics, University of Washington\\\\
  Shane Lubold \hspace{.2cm}\\
  US Census Bureau \\\\
  Arun G. Chandrasekhar \hspace{.2cm}\\
  Department of Economics, Stanford University, J-PAL, NBER \\\\
  Tyler H. McCormick\footnote{Correspondence: tylermc@uw.edu. The authors greatly thank Lori Beaman, Stephane Bonhomme, Vincent Boucher, Carlos Cinelli, Paul Goldsmith-Pinkham, Kosuke Imai, Ben Letham, Fabrizia Mealli, Alex Philip and Alex Volfovsky for helpful comments and discussion.} \hspace{.2cm} \\
  Departments of Sociology and Statistics, University of Washington
  }
  \date{\vspace{-50pt}}
   \maketitle
} \fi

\if0\blind
{
  \bigskip
  \bigskip
  \bigskip
  \begin{center}
    {\LARGE\bf Model-Based Inference and Experimental Design for Interference Using Partial Network Data}
\end{center}
  \medskip
} \fi

\bigskip
\begin{abstract}
The stable unit treatment value assumption states that the outcome of an individual is not affected by the treatment statuses of others, however in many real world applications, treatments can have an effect on many others beyond the immediately treated. Interference can generically be thought of as mediated through some network structure. In many empirically relevant situations however, complete network data (required to adjust for these spillover effects) are too costly or logistically infeasible to collect. Partially or indirectly observed network data (e.g., subsamples, aggregated relational data (ARD), egocentric sampling, or respondent-driven sampling) reduce the logistical and financial burden of collecting network data, but the statistical properties of treatment effect adjustments from these design strategies are only beginning to be explored. In this paper, we present a framework for the estimation and inference of treatment effect adjustments using partial network data through the lens of structural causal models. We also illustrate procedures to assign treatments using only partial network data, with the goal of either minimizing estimator variance or optimally seeding. We derive single network asymptotic results applicable to a variety of choices for an underlying graph model. We validate our approach using simulated experiments on observed graphs with applications to information diffusion in India and Malawi. \end{abstract}

\noindent%
%{\it Keywords:}  Interference,  Experimentation, Missing-Data,Networks, Optimal seeding, Potential-Outcomes

\vfill

\newpage
\spacingset{1.9} % DON'T change the spacing!
\section{Introduction} \label{sec: Introduction}
Interference occurs when one individual's treatment status impacts others' outcomes. Interference, also known as ``spillover effects,'' appears in multiple scientific domains, including the study of infectious diseases \citep{hudgens2008toward, tchetgen2012causal}, 
studying peer influence \citep{manski1993identification, bramoulle2009identification, de2010identification, epple2011peer, goldsmith-pinkhami2013},
public policy \citep{malani2021effect, imai2021causal}, information diffusion \citep{banerjee2013diffusion, banerjee2019gossip}, technology adoption \citep{beaman2021can}, online platforms \citep{saveski2017detecting, pouget2018optimizing, pouget2019testing} and online marketplaces \citep{ha2020counterfactual, johari2022experimental}, among other domains.

Interference violates the stable unit treatment value assumption (SUTVA), which states that an individual's outcome is not impacted by the treatment status of their peers. When SUTVA is violated, each potential outcome, the counterfactual outcome under a given treatment assignment, could depend on all treatment assignments within the population. Valid inference for treatment effects under SUTVA violations is an active area of research, with solutions typically depending on a combination of exposure maps and structural causal models. Exposure maps categorize respondents according to their network characteristics and the vector of treatment statuses~\citep{Aronow2017EstimatingExperiment, chandrasekhar2023general}, while structural causal models identify specific pathways for influence between individuals~\citep{van2012causal, ogburn2022causal}.

Estimating causal effects under interference typically requires complete network data, which is expensive and onerous to collect or may not be available due to privacy constraints. Partially observed network data takes many forms: subgraph samples where a researcher observes the presence/absence for only a subset of possible connections, egocentric sampling using either specific links or aggregates, or network-based sampling methods such as snowball sampling or respondent-driven sampling. In each case, incomplete network information introduces miss-measurement in the exposure map.  A person may have treated peers, for example, but if links to those peers are not observed, the researcher will think their outcome is totally orthogonal to the treatment.

This paper introduces a framework for estimation and inference of causal effects under \textit{partial} network data arising from a single graph. Partial here means that we may observe some or no links or aggregate summaries of links, which we will formalize later. With such data, we recover multiple estimands including various conditional or average treatment effects. To do this, we define a broad class of structural causal models that are amenable to estimation using partial data. This class covers many empirically relevant schemas for interference, such as diffusion and its generalizations.
Estimation leverages a dual approach: first, by using an iterated expectation method for de-biased estimation of model parameters with partial network data, and second, by managing the dependence of exogenous noise in the outcomes. ~\citet{chandrasekhar2011econometrics} also introduced an iterated expectations strategy for cases where multiple networks are available and data are independent across networks. We tackle the more challenging inference task of single network asymptotics.  Our method applies when the underlying graph has features captured by the class of node-exchangeable formation models, which we commonly see in practice and connect this to the problem of estimating effects of experiments.  Previous methods~\citep{breza2017using,breza2023consistently} developed a related method to estimate network model features using a specific type of aggregated network data.  Along with expanding to a wide range of partial data types, we extend this existing methodology to relax the requirement in previously published studies that traits be mutually distinct, a challenge to its usability in practice until now.

We also tackle the problem of experimental design associated with network exposure in scenarios where obtaining pristine network data ahead of randomized controlled trials (RCTs) is challenging or impossible. By collecting partial network data and employing a Bayesian optimization algorithm, we propose experimental designs that efficiently maximize treatment saturation tailored to specific estimands of interest. Our results demonstrate that this methodology not only surpasses traditional methods like inverse probability weighted (IPW) estimators in estimating global average treatment effects but also facilitates innovative seeding strategies that leverage the unique characteristics of partial network data.  We demonstrate that these techniques can be used to assign treatment in such a way as to minimize estimator variance or to optimally seed for diffusion. 

The remainder of the paper is structured as follows. We begin with a review of related work (section~\ref{sec:related}. section~\ref{sec: environment} defines the necessary background, then section~\ref{sec: Inference} describes the procedure for estimation and inference. section~\ref{sec: experimental design} describes experimental design using partial network data and section~\ref{sec: Data Analysis} provides empirical examples. We conclude in section~\ref{sec:conclusions}. Code to replicate the results in the paper is available at \url{https://github.com/SteveJWR/ardexp}, and an R package is available from \url{https://github.com/SteveJWR/SBMNetReg}.

\subsection{Related Work}
\label{sec:related}
We first provide a brief review of literature related to inference with partial network data, then move to an overview of causal inference under interference. Complete network data collection can be prohibitively expensive and restricted by privacy concerns \citep{breza2017using}. Researchers typically work with partial network data derived from various sources such as survey samples, coarse geographic data, kinship information from censuses, or aggregated financial transactions. Comprehensive reviews of methods for handling network data can be found in \citet{de2017econometrics} or \citet{graham2020network}, and discussions on identification in network and related models are provided in \citet{manski2009identification}. 
A direct approach is node subsampling, selecting a portion of nodes from the population and mapping the entire graph among them. If random sampling of nodes is infeasible, or if populations are sensitive or stigmatized, techniques like snowball or respondent-driven sampling offer a limited but focused view of the graph \citep{heckathorn1997respondent, goel2009respondent, goel2010assessing, baraff2016estimating, green2020consistency}

When complete edge enumeration among node subsets is impractical, researchers adopt standard survey methods such as Aggregated Relational Data (ARD) collection. The main intuition is that each of the partial network designs mentioned above can be used to estimate a breakdown of each respondent's network in terms of observable characteristics. In ARD surveys, respondents are asked, ``How many people do you know with trait X?" for various traits. Additional conditions may be added in addition to collect the type of connection that is relevant \citep{feehan2016quantity}.Originally designed to estimate hard-to-reach populations like HIV-positive men in the US \citep{killworth1998estimation, scutelniciuc2012network, jing2014estimating}, has been extended to a variety of other settings such as financial contagion models \citep{acemoglu2015systemic} as well as more general network scale up methods utilized (NSUM) \citep{killworth1998social,kadushin2006scale,feehan2016generalizing,mccormick2020network} and is notably 70 to 80\% less costly than full network data collection \citep{breza2017using}. %It proves valuable in numerous network inference challenges \citep{breza2017using, breza2023consistently}. 
Another standard survey method, egocentric sampling, asks respondents to consider specific individuals in their networks and provide detailed information about them, unlike the aggregate focus of ARD and is commonly used in applications such as contact tracing \citep{potter2011estimating}, violence perpetration \citep{bond2017contagious} or adolescent substance measurement \citep{huang2014peer}. 

The first task in causal inference problems, particularly in the presence of interference, is defining the target estimand.  The global average treatment effect (GATE), for example, assesses the impact of treating everyone versus treating no one, considering peer effects \citep{Ugander2013GraphUniverses}. Other interests might include the effect of specific treatment allocations, like identifying influential individuals \citep{kempe2003maximizing, banerjee2019gossip}, often limited by policy constraints (e.g., subsidies for the ultra-poor as in \citet{anderson2007agricultural}) or due to non-monotone peer effects dynamics \citep{banerjee2018less}: treating everyone may change interaction dynamics in equilibrium. More generally \citet{Aronow2017EstimatingExperiment} compare average treatment effects between two exposure configurations. A distinct but related line of work seeks to detect whether interference is present at all \citep{athey2018exact}. 

Models for peer influence like contagion \citep{jackson2008social, banerjee2013diffusion, beaman2021can, xiaoqi2023measuring} or hearing models \citep{banerjee2019gossip} structure interference analysis by identifying specific mechanisms that describe how connections between peers impact outcomes. \citet{auerbach2021local} explore these effects through structural causal models focusing on nonparametric estimation, while our work emphasizes estimation, inference, and design using partial network data.  Much of this literature assumes a fully observed graph, though a recent line of literature address imperfect or incompletely sampled graphs under certain conditions and for specific average causal effects \citep{Hardy2019EstimatingNetworks, yu2022estimating, cortez2022staggered}. A related line of work examines sensitivity analysis for standard causal estimators under hidden treatment diffusion \citet{tortu2021causal}.

\section{Environment} \label{sec: environment}

%\subsection{Data} 
Let $i \in \{1,2,\dots, n\} = \mathscr{V}$ denote a population of interacting individuals and let $\mathscr{G} = \mathscr{V} \times \mathscr{E}$ be the network by which interference is propagated; where $\mathscr{V}$ is the set of node vertices and $\mathscr{E} \subset \mathscr{V} \times \mathscr{V}$ is a set of edges (either directed or undirected). We can also extend this to weighted graphs, however binary networks are presented for simplicity. We can represent this graph by the adjacency matrix $G \in \{0,1\}^{n \times n}$. We consider binary treatments denoted by a treatment vector $\vect{a} \in \{0,1\}^n$ and let denote the potential outcome $Y_i(\vect{a}) \in \mathbb{R}$, under a treatment assignment $\vect{a}$, and $Y_i$ denote the actual observed outcome. Lastly, we assume that we have access to pre-treatment node-level covariates $X_i \in \R^{m}$.
In the remainder of the paper let $O$ and $o$ denote the usual big and little oh notation and $O_P$ and $o_P$ denote the stochastically bounded and convergence to $0$ in probability for sequences of random variables. We use $\tilde{O}$ if we are suppressing logarithmic factors in the rate. Let $\norm{\cdot}_{p}$ denote and $p$-norm, and let $\norm{\cdot}_F$ denote the Frobenius norm. 

\subsection{A structural causal model} \label{sec: SCM}
We use the framework of structural causal models, a nonparametric extension of structural equation models \citep{pearl2009causality}. Similar approaches have been studied by \citet{ogburn2022causal} and \citet{auerbach2021local} in the case of fully observed networks.  We derive a model that is amenable to estimation with partial data. 

Let $Y_i(\vect{a})$ denote the potential outcome of $Y_i$ under a treatment allocation $\vect{a}$. The exposure mapping $V_i$ is represented as a function $f_V$ such that $V_i = f_{V}(\vect{a}, \varphi_i(G)) \in \R^{p_V}$ where $\varphi_i$ is the relevant graph information for individual $i$ relative to their position with respect to treated individuals. We also allow for the potential outcome to be modulated by some additional confounder $S_i = f_{S}(\vect{X}, \vartheta_i(G)) \in \R^{p_S}$. We model the potential outcomes $Y_i$ as a function of the exposure, type-value $S_i$ and some additional noise $\vect{\varepsilon}_{Y}$ 
\begin{equation} \label{eq: SCM}
        Y_i = f_Y(S_i, V_i, \vect{\varepsilon}_{Y})
\end{equation}
The benefits of structural causal models are that they allow for the characterization of all causal effects in a system, as well as the distributions of counterfactuals. However, they require correct specification of the causal process, i.e. correct specification of the exposure map and the relevant confounders. Even if one can propose a model for interference, estimation is not straightforward due to the fact that we only observe partial graph information in $G^*$. Many common models of interference can be expressed as structural causal models, and can be thought of as parameterizations of $f_Y(S_i, V_i, \vect{\varepsilon}_{Y}) = f_Y(S_i, V_i, \vect{\varepsilon}_{Y}; \beta_0)$. This then reduces the challenge to estimating $\beta_0$ using partially observed data. The exogenous noise, $\epsilon_Y$, within our model is likely influenced by the graph structure, as interactions and peer effects can induce correlations in outcomes that extend beyond individual exposures. This complexity suggests that the noise, even if initially considered as external to the model, is intertwined with the network dynamics, reflecting the propagation and interference effects inherent in our structural causal framework. 

We distinguish two types of target parameters. The first are the \textbf{outcome model} parameters, which parameterize the distribution of the outcome, exposure, and confounder $(Y,S,V)$. Specifically, $f_Y(S_i, V_i, \epsilon_Y) = f_Y(S_i, V_i, \epsilon_Y;\beta_0)$ under parameterization $\beta \in \mathbb{R}^p$. The true model parameters are $\beta_0 \in \R^p$, identifiable through a moment equation $m$, $\E[m(Y_i,S_i,V_i, \beta_0)] = 0$, or through regression. In a simple diffusion model, this is the probability of infecting a neighboring node $q \in [0,1]$.

The second set of parameters we consider are the \textbf{causal} parameters, those involving the distributions of the counterfactuals. The main causal parameter we will consider is the expected average potential outcome on the complete network $G$, $\Psi(\vect{a}|G) = \frac{1}{n}\sum_{i = 1}^n\mathbb{E}[Y_i(\vect{a})]$, though these can also be made conditional on a covariate $x$: $\Psi(\vect{a}|x, G)$. Leveraging the structural causal model, we can define the these causal effects in terms of the structural causal model. We illustrate conditions for identification of these causal effects in section~\ref{sec: nonparamtericID}. While we focus on defining causal quantities through conditional means, the nonparametric identification can also apply to other functionals like quantiles. 

Inference for learning the causal parameters under the above assumptions now amounts to learning the distributional relationship between $Y_i$ and $S_i,V_i$. We consider settings where the assignment of treatments can be manipulated by an experimenter, which we discuss in section~\ref{sec: experimental design}. If one leverages this model, either through assumption or estimation, then we can use a structural causal model to generate expected potential outcomes under different treatment assignments $f_Y(S_i, V_i, \vect{\varepsilon}_{Y})$, which is precisely what is done in the case of seeding. A contrast of these frameworks is included in the Appendix in section~\ref{sec: causal frameworks}. The applicability of a model to a new population parallels challenges in distribution shift, as explored in \citet{shimodaira2000improving} or \citet{wilkins2024multiply}.

Adding structure to the potential outcomes model is standard in fields like economics, where researchers often propose models to explain how information or behaviors spread across networks. Many of these models include a temporal element. In our setting, we consider outcomes at a fixed time $T$, i.e., $Y_i(\vect{a}) = Y_{i,T}(\vect{a})$.  For instance, \citet{banerjee2013diffusion} explore a latent diffusion process in micro-lending, \citet{banerjee2019gossip} study a hearing model for information diffusion, and \citet{beaman2021can} analyze behavior adoption in agriculture through complex contagion. Additionally, \citet{centola2007complex} differentiate the spread of information, often through single links, from behaviors that require multiple neighbors for network propagation. 

\subsubsection{Example: Contagion as a structural causal model} \label{sec: example contagion}
A foundational model of information diffusion is based on simple contagion, and generalizations of SIR (Susceptible-Infected-Recovered) models on networks \citep{kermack1927contribution, giles1977mathematical}. These models have been further studied and extended in various settings \citep{jackson2006diffusion, aral2009distinguishing, romero2011differences, chierichetti2011reconstructing, banerjee2013diffusion, banerjee2019gossip}.
Here we illustrate how the base model, under which many extensions are built, can be interpreted as a structural causal model. This interpretation can also be applied to complex contagion settings \citet{centola2007complex, beaman2021can, cencetti2023distinguishing}.

\begin{figure}[htbp]
\centering
\subfigure[Base Network]{
\includegraphics[width = 0.25\textwidth,trim={10cm 3cm 7cm 3cm},clip]{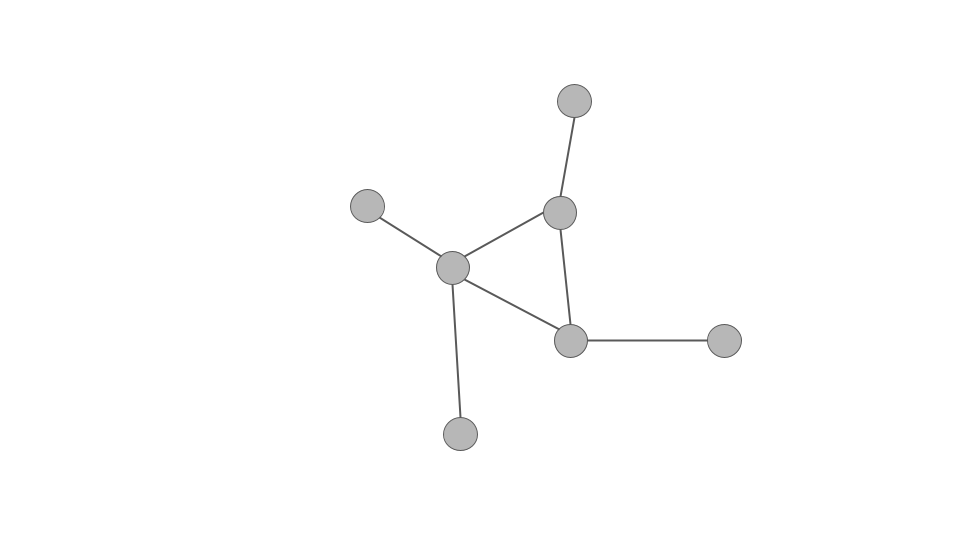}
}
\subfigure[Noise draw $\epsilon_{ij}$]{
\includegraphics[width = 0.25\textwidth,trim={10cm 3cm 7cm 3cm},clip]{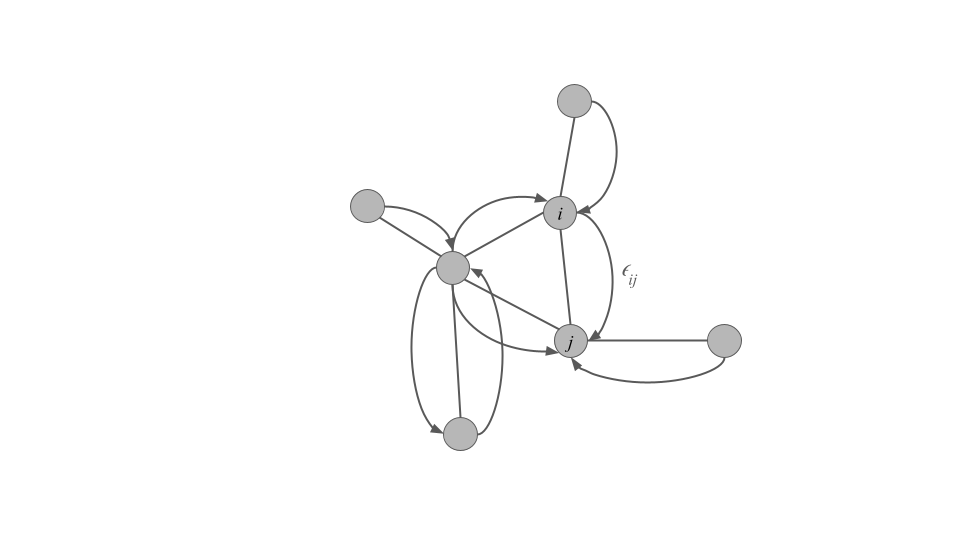}
}
\subfigure[Directed Network $\vect{D}$.]{
\includegraphics[width = 0.25\textwidth,trim={10cm 3cm 7cm 3cm},clip]{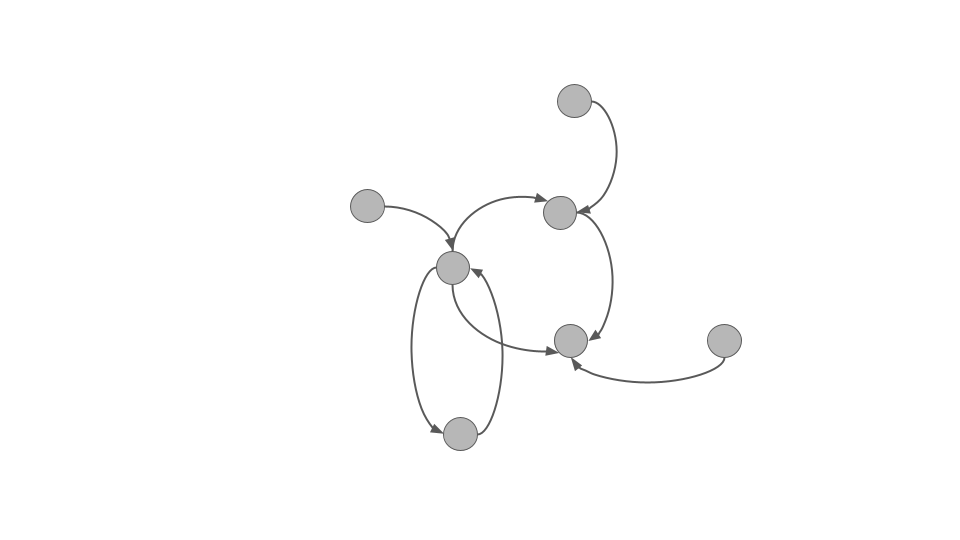}
}
\\ 
\centering
\subfigure[Contagion process at time period $T = 0$.]{
\includegraphics[width = 0.25\textwidth,trim={10cm 3cm 7cm 3cm},clip]{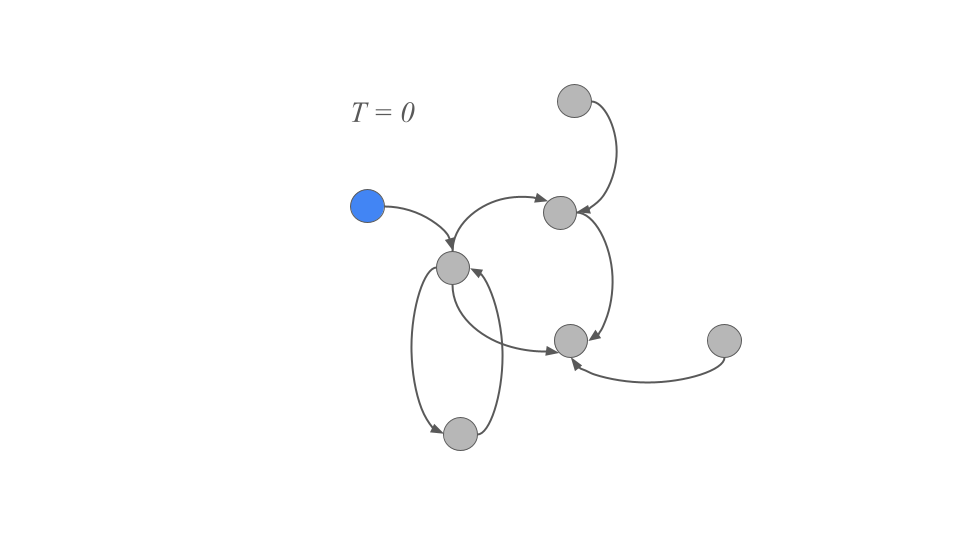}
}
\subfigure[Time period $T = 1$.]{
\includegraphics[width = 0.25\textwidth,trim={10cm 3cm 7cm 3cm},clip]{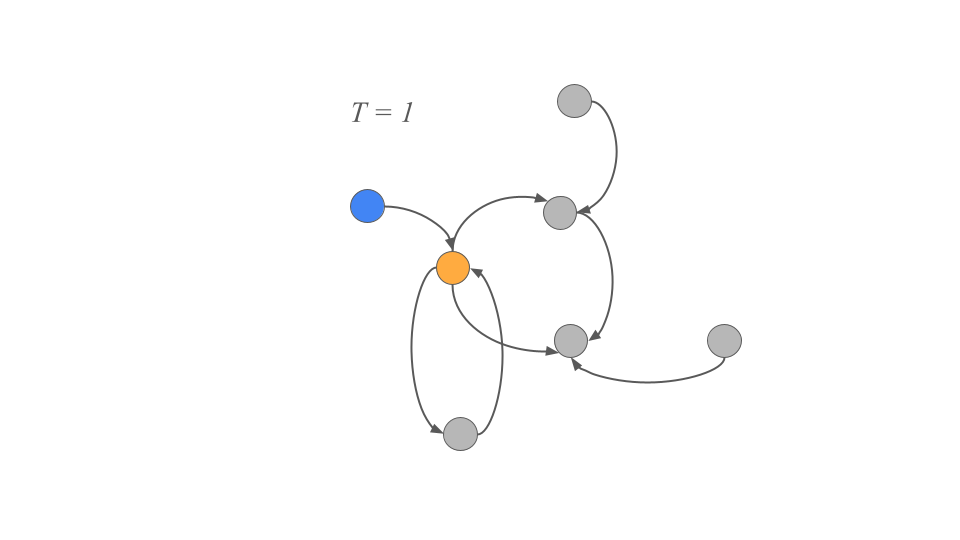}
}
\subfigure[Time period $T = 2$.]{
\includegraphics[width = 0.25\textwidth,trim={10cm 3cm 7cm 3cm},clip]{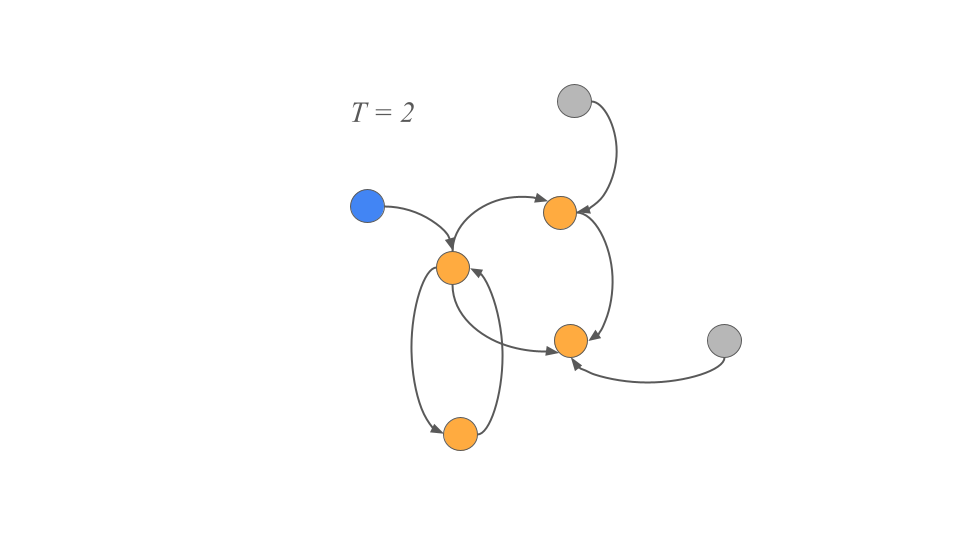}
}
\caption{Contagion process where a single node is seeded in time $T = 0$ in blue, and infected nodes displayed in orange at times $T = 1$ and $T = 2$.}
\label{fig: Diffusion Example}
\end{figure}

Consider a scenario where initially infected (treated) seeds $\vect{a}$ transmit the infection to each neighbor with probability $q$ at each time-step $t \in {1,2,\dots, T}$, after which they are no longer infectious. An infection status at time $t$ is denoted as $Y_{it} = 1$. The overall outcome $Y_i= 1$ indicates whether a node was infected at any time up to $T$. For a simple case with $T = 2$, we model the transmission using Bernoulli random variables $\epsilon_{ij} \sim \text{Bernoulli}(q)$, representing potential infection from node $i$ to node $j$. Let $\vect{E}_{ij} = \epsilon_{ij}$ and $\vect{D} = \vect{E} \odot G$. This setup is depicted in Figure~\ref{fig: Diffusion Example}.
Given a random sample of the directed graph $\vect{D}$, one can characterize what would have happened if a node were treated, which is precisely the counterfactual. For instance in Figure~\ref{fig: Diffusion Example} we seed the left most which proceeds to propagate in steps $1$ and $2$. Additionally, one can construct the  relevant exposure map for any fixed number of time steps $T$.

\subsubsection{Examples of Exposure Maps} \label{sec: examples of exposure maps}
We consider several examples of exposure maps, though this list is not exhaustive. 

\begin{example}[Local Interaction Effects]
    Simple examples of local network effects may include the total number of treated neighbors,
$V_i = \sum_j G_{ij}a_j,$
or the treated fraction of one's neighbors $V_i = \sum_j \frac{G_{ij}}{d_i}a_j$, 
where $d_i = \sum_{j} G_{ij}$ is degree.

\end{example}

\begin{example}[Risk-Sharing Networks \citep{ambrus2014consumption}]

Equilibrium risk sharing is that the graph consists of $C$ mutually exclusive communities such that any endowment vector within the community is aggregated and shared evenly. Let treatment $\vect{a}$ be an ``endowment'' and let $\bar{ \vect{a}}_c = \sum_{j \in c} a_j$ be the sum of the endowment vector for community $c$, with $|c|$ denoting its size. Then,
$V_i = f_V(\vect{a},\varphi_i) = \bar{ \vect{a}_c} \cdot |c|^{-1}.$
That is, the exposure is just a function of the total endowment of the community and nothing more.
\end{example}
\begin{example}[Hearing Information \citep{banerjee2019gossip}] \label{example: hearing model exposure map}
Many phenomena, like the spread of diseases, information, or social behaviors, can be effectively modeled as contagion processes. These models show how such phenomena spread through networks \citep{keeling2008modeling,centola2007complex,barrat2008dynamical,pastor2015epidemic,cencetti2023distinguishing}.

\citet{banerjee2019gossip} introduces a message-passing model based on such a contagion process. The treatments, denoted by $\vect{a}$, represent a seed piece of information disseminated over a series of time steps, from 1 to $T$. After $T$ time steps, no further message spreading occurs. We define a ``hearing matrix" $\vect{H}_0$, which calculates the expected number of times person $j$ hears information from person $i$ after $T$ time steps, based on transmission probabilities.

The expected total number of messages that person $j$ hears by time $T$ is represented by $V_j$ (the exposure) which affects their response $Y_i$ through a link function $\Lambda$:
\begin{equation*}
    \E[Y_i|V_i] = \Lambda(\beta_0 + \beta_1 V_i).
\end{equation*}
A common assumption is propose a single transmission probability for each individual $q$, thus giving structure to the exposure map:
\al{
    V_i &=  (\vect{H}\vect{a})_{i} \text{ the $i^{th}$ element of this vector} \\
    \text{where } \vect{H} &= \sum_{t = 1}^T q^t G^{t}
}
It is straightforward to generalize this to include heterogeneity in the diffusion time steps $\beta_t$ and illustrate this model in equation~\eqref{eq: heterogeneous diffusion}:
\begin{equation}
    \E[Y_i| V_i]  = \Lambda(\beta_0 + \sum_{t = 1}^T \beta_t \vect{e}^T_i G^{t} \vect{a}) = h(S_i, V_i; \beta) \label{eq: heterogeneous diffusion}.
\end{equation}
Furthermore, we can relax this model to allow for additional heterogeneity through graph-level statistics $S_i$, which may include node-level covariates $X_i$ or individual graph-level information such as the degree $d_i$.

\end{example}

\subsection{Examples of Partially Measured Network Data}
Now that we've established our framework for interference, we next return to the data used for estimation.  In our setting, we do not have access to the full graph $G$, but rather, have access to some summarizing function the graph $G^* = \zeta(G)$. \citet{tsiatis2006semiparametric} uses the term coarsened data to refer to such partial measurements of missing data in general, not necessarily in the network setting. ``Coarsened'' is apt in our setting because our method works on partial graph structure that give (either directly or by construction) an estimate of linking rates across population members with different combinations of observable traits.  A non-exhaustive set of examples of partially measured network data include induced subgraphs or egocentric sampling \citep{freeman1982centered, almquist2012random}, respondent driven sampling \citep{heckathorn1997respondent}, aggregated relational data \citep{killworth1998estimation}, respondent driven sampling \citep{heckathorn1997respondent, goel2009respondent, goel2010assessing, green2020consistency} and more.   

\begin{example}[Induced subgraph]
    We sample $m \leq n$ of nodes in the graph randomly, with at least one node from each of the $K$ communities. Let $G^* = G_{I_m,I_m}$ be the sub-graph induced by these $m$ nodes where $I_m \subset \{1,2,\dots, n\}$ are the set of nodes that are sub-sampled from the whole population. 
\end{example}

\begin{example}[Respondent Driven Sampling]
    Let $i \in I_m \subset \{1,2,\dots, n\}$ denote the indices of a sample of individuals obtained through respondent driven sampling. An initial number of individuals are recruited as seeds, and subsequent individuals are recruited via referrals from the others in a population. Under this process we receive a subgraph of connected individuals $G_{I_m,I_m}$ as well as the list of connections to additional nodes $I_{n\setminus m} = \{1,2,\dots, n\} \setminus I_{m}$ $G^* = G_{I_m,I_m}, G_{I_m,I_{n\setminus m}}$. 
\end{example}

\begin{example}[Aggregated Relational Data]\label{exp: ard_example}
    Aggregated relational data consists of aggregated sums of connections to nodes of a given trait. Typically this is collected from a survey consisting of questions such as ``How many many people do you know with [X] trait?''. For a set of $T$ traits, this consists of
        $X^*_{it} = \sum_{i = 1}^n G_{ij}I(t_j = t)$.
\end{example}

In order to infer about the distribution of the missing part of the graph, we propose that $G \sim \theta_0$ where we assume that $\theta_0 \in \Theta$ denotes the parameters of a random graph model. In this case, for each $i$, there is an a latent $\xi_i$ parameter such that  $P(G_{ij} = 1 |\xi_i, \xi_j) = \tilde g(\xi_i, \xi_j)$ for some function symmetric, measurable, $\tilde g$, known as a graphon \citep{lovasz2006limits, orbanz2015bayesian}. Many common graph models, such as latent space models \citep{Hoff2002LatentAnalysis, handcock2007model, lubold2023identifying, wilkins2022asymptotically}, are included in this category. Graphons are appealing in this context because, following~\citet{airoldi2013stochastic} and ~\citet{gao2015rate}, they can be approximated arbitrarily well using latent types assigned to each node. Said another way, graphons introduce complex dependence in the network-generating mechanism through clustering induced by latent types associated with each node. 
In our inferential procedures in section~\ref{sec: Inference}, the general procedures involve estimation from a missing data perspective. This will involve estimating the graph model $\hat \theta := \hat \theta(G^*)$ then inferring about the distribution $G|G^*, \hat \theta$. Further details for estimating the graph model are included in section~\ref{sec: Network Model Estimation}.

\subsection{Nonparametric Identification of Causal Effects} \label{sec: nonparamtericID}

We first illustrate an identification procedure for the causal effect without a-priori imposing any model structure. These are analogous to standard causal identification assumptions, adapted to our framework. 
\begin{definition}[Exposure Weak Ignorability]
    We say that an exposure assignment is \textbf{weakly ignorable} if the following holds: 
    $$ Y_i(v) \indep \{V_i =v\} | S_i$$
\end{definition}

Conditioning the graph confounder $S_i$ captures the heterogeneity of the outcomes when observed with a given exposure. In simple contagion models, nodes are equivalent, and this independence occurs naturally without conditioning. section~\ref{sec: causal effect estimation} discusses an example from \citet{ugander2023randomized} where conditioning on node degree suffices for any randomization.
\begin{definition}[Exposure Consistency]
    Exposure consistency holds if 
    $$ V_i = v \implies Y_i = Y_i(v) = Y_i(\vect{a})$$
    where $Y_i(v)$ is the potential outcome of individual $i$ for the exposure $v$. 
\end{definition}
This assumption can be simply understood as the exposure is correctly specified.
\begin{definition}[Conditional Independence of the Graph and Outcome]
    We assume that the outcome is conditionally independent of the outcome conditional on the exposure and the graph generative parameters 
    $$ Y_i(\vect{a}) \indep G | V_i, S_i. $$
\end{definition}
This assumption states that once we have adjusted for $V_i$ and $S_i$, then the potential outcomes are independent of the network $G$. These assumptions allow us to express the causal estimand through observational data. 
\begin{align*}
    P(Y_i(\vect{a}) = y|S_i = s, G) &= P(Y_i(v) = y|S_i = s, G) \text{ By the exposure mapping} \\
    &= P(Y_i(v) = y| V_i = v, S_i = s, G) \text{ By weak ignorability} \\
    &= P(Y_i = y| V_i = v, S_i = s, G) \text{ By consistency} \\
    &= P(Y_i = y|V_i = v, S_i = s) \text{ Graph conditional independence} \\
    \implies \Psi(\vect{a}| G) &= \frac{1}{n} \sum_{i = 1}^n \mathbb{E}[Y_i|V_i = f_{V}(\vect{a}, \varphi_i), S_i = f_{S}(\vect{X}, \vartheta_i)]
\end{align*}
For brevity, we denote the true conditional mean $\mathbb{E}[Y_i|V_i = v, S_i = s]$ as $h_0(s,v)$ and denote $h(s,v;\beta)$ a model to estimate $h_0(s,v)$. Given a network model $\theta$, observed graph data $G^*$, and a conditional model $h(s,v;\beta)$ we can also define the expected average treatment effect 
\begin{equation}
    \Psi(\vect{a}| \beta, G^*, \theta) = \frac{1}{n} \sum_{i = 1}^n \E[h(f_V(\vect{a},\vect{X};\varphi_i),f_{S}(\vect{X}; \vartheta_i);\beta)| \vect{a}, \vect{X}, G^*, \theta] 
\end{equation} 
where under the correct model conditional model and graph model $\E[\Psi(\vect{a}| G)|\vect{a}, \vect{X}, \theta_0] = \Psi(\vect{a}| \beta_0, G^*, \theta_0)$. In Appendix~\ref{sec: plug-in estimates}, we illustrate when this population average effect under any draw of the network $\Psi(\vect{a}| G)$ will be close to the average over the model class $\Psi(\vect{a}| \beta_0, G^*, \theta_0)$; and study plug-in estimators $\hat \Psi(\vect{a}| G) =  \Psi(\vect{a}| \hat \beta, G^*, \hat \theta)$. 

\section{Inference} \label{sec: Inference}

We outline our method for estimating parameters with partial network data. Developing these results requires two theoretical tools: a fast estimation rate for network model parameters $\theta_0$, and a suitable central limit theorem for scenarios with correlated outcomes.
Outcome Model Parameters and Estimators
Next we consider estimating the outcome model parameters $\beta_0$. We present two methods for estimating such parameters, instrumentation in a linear model, and $Z$ estimators. The iterated expectation procedure for estimating such parameters was introduced in \citet{chandrasekhar2011econometrics}, however, we extend inference to the single network setting. Similar approaches exist for peer effects models \citep{boucher2020estimating}. 

\subsubsection{Estimation in Linear Models}
We first illustrate identification of the conditional model under a linear model assumption. 
\begin{equation*}
    Y_i = \beta_0^T \tilde h(S_i,V_i) + \varepsilon_i
\end{equation*}
where $\E[\varepsilon_i] = 0$ and there can be general correlation $\Var[\vect{\varepsilon}] = \Sigma$. Without access to the network data, one can recover the model parameters through conditional expectation
\al{
    \E[\E[Y|S(G),V(G),G,\vect{a}, \vect{X}]| \vect{a}, \vect{X}, G^*, \theta_0] &= \beta^T_0 \E[\tilde h(S(G), V(\vect{a}, G))| \vect{a}, \vect{X}, G^*, \theta_0]
}
where we create a new set of features $\tilde H_i= \E[\tilde h(S_i(G), V_i(\vect{a}, G))| \vect{a}, \vect{X}, G^*, \theta_0]$ by averaging over the network model. Here identification comes from the variation of these averaged features $\tilde H_i$ over the population. More clearly, letting $\vect{\tilde H} \in \R^{n \times p}$ denote the design matrix of this model, identification comes from the linear independence of the columns of $\vect{\tilde H}$. 

\subsubsection{Z estimators}
In other cases, parameters may be defined through a moment equation, and can be used to construct a $Z$-estimator, for example, generalized linear models (GLMs). These parameters can be identified using an estimating equation approach where given a moment function $\tilde m(Y_i,S_i,V_i; \beta)$ such that $\E[\tilde m(Y_i,S_i,V_i; \beta)|\vect{a}, \vect{X}, G] = 0$ if and only if $\beta = \beta_0$. Through the use of iterated expectations, we can define a new estimating equation, by marginalizing over the draws of the graph model then applying iterated expectations
\begin{align}
    m_i(Y_i; \beta, \vect{a}, \vect{X}, G^*, \theta) &:= \E[\tilde m(Y_i, S_i, V_i; \beta) |Y_i, \vect{a}, \vect{X}, G^*, \theta ] \label{eq: observed data estimating equation} \\
    \text{where }\E[m_i(Y_i,S_i,V_i; \beta_0, \theta_0)| G^*, \vect{a}, \vect{X}, \theta_0] &= \E[\E[\tilde{m}(Y_i,S_i,V_i; \beta_0)|\vect{a}, \vect{X}, G]| G^*, \vect{a}, \vect{X}, \theta_0] = 0. \nonumber
\end{align}
Identification arises from the variation of exposure and confounders, such that  $\beta = \beta_0$  if and only if $\E\left[\frac{1}{n}\sum_{i = 1}^n m_i(Y_i, S_i, V_i; \beta, \theta_0)\bigg| G^*, \vect{a}, \vect{X}, \theta_0\right] = 0$. Exact conditions vary by parameter, but GLMs can use a similar strategy as linear models.

\subsection{Inference with partially measured data.} \label{sec: partially measured data inference}
We introduce a general procedure for estimating the outcome model parameters. We also illustrate inference for estimation of a causal target parameter on a particular graph $G$. We present a pseudo-code approach to the procedure in Algorithm~\ref{alg: missing data z-estimation}. Let $\tilde Z_i= (Y_i, S_i, V_i)$ denote the full (including unobserved) data, and let $\vect{Z} = (\vect{Y}, \vect{a}, \vect{X}, G^*)$ denote the observed data.

\begin{algorithm} 
\footnotesize
\caption{Z-estimation overview}
\begin{algorithmic}[1]
    \State Define an model for the relationship of $\vect{Y}$ given the exposures $\vect{V}$ and confounders $\vect{S}$ (for instance, a regression model $\E[Y|V,S] = h(v,s;\beta_0), \beta \in {\mathcal{B}} \subset \R^p$ with parameters which can be estimation via the estimating function $\tilde m_n(\vect{\tilde Z}; \beta)$. Let $\tilde m_n(\vect{\tilde Z}; \beta) = \frac{1}{n}\sum_{i=1}^n m(\tilde Z_i; \beta)$ denote the empirical estimating function. \label{alg_step: regression outcome}
    \State Estimate a model of the network, using the node-level covariates $\hat \theta := \hat \theta(G^*)$. \label{alg_step: estimate blockmodel}
    \State Estimate $\hat \beta$ by solving the estimating equation $m_n(\vect{Y}; \hat \beta, G^*, \widehat{\theta}) =0$, where $m_n(\vect{Y}; \hat \beta, G^*, \widehat{\theta}) = \frac{1}{n}\sum_{i = 1}^n m_i(Y_i; \beta, \vect{a}, \vect{X}, G^*, \theta)$ where $m_i$ is defined in equation~\eqref{eq: observed data estimating equation}. \label{alg_step: marginal solution}
    \State (optional) Plug in $\hat \beta$ to $\Psi(\vect{a}| \hat \beta, G^*, \hat \theta)$. \label{alg_step: causal plug-in}
\end{algorithmic}\label{alg: missing data z-estimation}
\end{algorithm}

Step \ref{alg_step: regression outcome} asks the practitioner to propose a response model given the treatment, i.e. the causal model in section~\ref{sec: SCM}. Step~\ref{alg_step: estimate blockmodel} estimates the generative model given the partial network data and the node covariates observed.  We give theoretical results where the formation model is a stochastic blockmodel, then give rate estimation relative to the more general graphon approach.  Step~\ref{alg_step: marginal solution} estimates the parameter by marginalizing the estimating function over the graph model. Lastly, Step~\ref{alg_step: causal plug-in} is optional if the target parameter is a plug-in estimator of the causal parameter using the regression model. We discuss inference for the plug-in estimate of causal parameters using a delta method argument in the Appendix~\ref{sec: plug-in estimates}. We next give our asymptotic results, then provide an example of this algorithm in section~\ref{sec: causal effect estimation}.

\subsection{Asymptotic Results}

The asymptotic results for both the Z-estimator and the linear model will depend on being able to establish a central limit theorem based on the exogenous noise. To establish asymptotic properties for our outcomes on a network, we extend the application of the central limit theorem (CLT) to structures not commonly associated with traditional time series or spatial dependencies. Nonetheless when the exogenous noise is correlated, we will need a method of handling the central limit theorem. Specifically, we utilize a general version of the CLT for dependent data from \citet{chandrasekhar2023general}. For brevity in presentation, we leave the full detail of this central limit theorem to the appendix. 

We denote $g_i(\vect{Z}; \beta) = m_i(\vect{Y}; \vect{a}, \vect{X}, \beta, G^*,\theta_0)$ to be the moment function evaluated using the true generative model and correspondingly $g_n(\vect{Z}; \beta) = \frac{1}{n}\sum_{i = 1}^n g_i(\vect{Z}; \beta)$. Further, define the (normalized) random vector of the estimating function evaluated at the correct model parameters $\mathcal{E}_i = \frac{1}{n}g_i(\vect{Z}; \beta_0)$. And lastly let $D_n(\vect{Z}; \beta_0) = \nabla_{\beta} g_n(\vect{Z}; \beta_0) \in \R^{p \times p}$ denote the gradient of the estimating equation $g_n(\vect{Z}; \beta)$. 
To develop valid inference, we must estimate the graph model quickly enough to disregard the graph estimation component during inference. We will next present the theorem and discuss the assumptions further.

\begin{assumption}[Regularity Conditions for Z-Estimation] \label{assumption: Z estimator regularity conditions}
    Suppose the following conditions hold for all $n$. \\
    \textbf{Consistency for a Z estimator}
    \begin{enumerate}[label=A\arabic*.,ref=A\arabic*]
    \item  $\mathbb{E}[g_n(\vect{Z};\beta)] = 0$ for $\beta = \beta_0$ and for all $\epsilon > 0$, $\text{inf}_{\norm{\beta - \beta_0} > \epsilon} \mathbb{E}[g_n(\vect{Z};\beta)] > 0$ \label{assumption: separatedness of solution}
    \item $ \sup_{\beta \in {\mathcal{B}}}\bigg| \left( \frac{\partial}{\partial \beta} \right)^l g_n(\vect{Z};\beta) - \left( \frac{\partial}{\partial \beta} \right)^l \E[g_n(\vect{Z};\beta)]\bigg|= o_P(1)$ for $l \in \{0,1,2\}$  \label{assumption: Uniform LLN}

    \end{enumerate}
    \textbf{Graph Model Regularity conditions}
    \begin{enumerate}[label=B\arabic*.,ref=B\arabic*]
    \item $\hat \theta$ is an $s(n)$-consistent estimate of the graph parameters $\norm{\hat \theta - \theta_0} = o_P(s(n))$ \label{assumption: Model Estimation Rate}
    \item $\sup_{\beta \in {\mathcal{B}}} |m_n(\vect{Z}; \beta, \theta) - m_n(\vect{Z}; \beta, \theta')| \leq b_n(\vect{Z}) \norm{\theta - \theta'}$ where $b_n(\vect{Z}) = O_P(1)$ (that is, $b_n(\vect{Z})$ is stochastically bounded). 
    \end{enumerate}
    \textbf{Central Limit Theorem (CLT)}
    \begin{enumerate}[label=C\arabic*.,ref=C\arabic*]
    \item The random vectors $\mathcal{E}_{1:n}$ satisfy the affinity set conditions of \citet{chandrasekhar2023general} (restated as Theorem~\ref{theorem: generalized CLT} in the appendix) with corresponding covariance matrix $\Gamma_n = \text{Var}\left[ \sum_{i = 1}^n \mathcal{E}_{i} \right]$. Where $r(n) := \sqrt{\lambda_{min}(\Gamma_n)}$. 
    \label{assumption: CLT of the estimating function} 
    \end{enumerate}
\end{assumption}
\begin{theorem}[Single Network Z-estimator Asymptotics] \label{theorem: Z estimator asymptotics}
    Suppose that $\hat \beta$ is computed as per Algorithm~\ref{alg: missing data z-estimation}, Assumptions~\ref{assumption: Z estimator regularity conditions} hold, and $s(n) = o(r(n))$. 
    Then: 
    \begin{equation}
        \Gamma_n^{-1/2}D(\beta_0)(\hat \beta - \beta) \to_d N(0, I_p)
    \end{equation}
    Where $\E[\nabla_{\beta} g_n(\vect{Z}; \beta)|\vect{a}, \vect{X}, G^*, \theta_0] \bigg|_{\beta = \beta_0} = D(\beta_0)$
\end{theorem}

The first set of assumptions ensures the consistency of $Z$-estimators, typically derived from uniform laws of large numbers as discussed in \citet{andrews1987consistency,newey1994large}. The second set involves conditions that make the graph model's estimation negligible, requiring the estimating functions to be smooth with respect to the graph parameters.

The final set of assumptions, stated in \ref{assumption: CLT of the estimating function}, are utilized so that $\mathcal{E}_i$ satisfy a central limit theorem \citet{chandrasekhar2023general}. This assumption is required if the data exhibit further dependence after controlling for graph parameters (if, for example, there are latent factors that impact both outcomes and the propensity to form ties).  The main idea of~\citet{chandrasekhar2023general} is to represent dependence in terms of ``affinity sets'' where the majority of dependence structure is captured within sets, leaving little between sets.  In the modelling of social behaviours beyond just considering outcomes as a function of the exposure observed, outcomes may be further correlated, beyond examples of spatial dependence or heteroskedasticity.  In practice we can include these dependencies through correlation terms matching the generative graph model, such as between blocks of a stochastic blockmodel or via latent positions in a latent space model.

Here, $r(n)$ describes the effective rate at which the variance converges. For the estimation of the graph model $\theta_0$ to be considered negligible, it must occur more rapidly than $r(n)$. In cases of independent or minimally dependent noise, it is typical for $r(n) \approx n^{-1/2}$. Alternatively, in different scenarios, $\mathcal{E}_i$ might exhibit correlation within densely connected blocks of the network, such as during a diffusion process in a stochastic blockmodel with $k_n$ densely linked blocks (refer to \citet{chandrasekhar2023general}, section 4.4). In such cases, $r(n)$ is generally on the order of $k_n^{-1/2}$. If both $r(n)$ and $s(n)$ approach zero, but $\frac{s(n)}{r(n)}$ diverges or stabilizes at a nonzero constant, a consistent estimator for the outcome model parameters can still be obtained. However, its asymptotic distribution may be influenced by the graph model estimation, necessitating a tailored inference approach.

An analogous argument follows when conducting inference using a linear model. For the sake of brevity and avoiding repetition, we include it in the Appendix in section~\ref{sec: OLS estimator inference}. In Theorem~\ref{theorem: ols asymptotics summary} we present a summary. 

\begin{theorem} \label{theorem: ols asymptotics summary}
    Let $\tilde H_i(\theta) = \E[\tilde h(S_i(G), V_i(G))|\vect{a}, \vect{X}, G^*; \theta]$. The OLS estimator uses the model averaged coefficients $\tilde H_i(\theta)$ in place of the true unobserved coefficients $\tilde h_i$. Let $\mathsf{H}_n(\theta) = \frac{1}{n} \sum_{i = 1}^n \tilde H_i(\theta) \tilde H^T_i(\theta)$. Given an estimate of the model parameters $\hat \theta$,  we define the 
    \[
        \hat \beta_{lm} = \mathsf{H}^{-1}_n(\hat \theta) \frac{1}{n} \sum_{i = 1}^n \tilde H_i(\hat \theta) Y_i 
    \]
    Let $u_i = (\tilde h(S_i(G), V_i(G)) - \tilde H_i(\theta_0))\beta_0 + \epsilon_i$ and let $\Gamma_n = \text{Var}\left[ \frac{1}{n}\sum_{i = 1}^n u_i \right]$  Suppose the conditions of Theorem~\ref{theorem: ols asymptotics} in the Appendix hold. 
    Then
    \[
        \Gamma_n^{-1/2} \mathsf{H}_n(\hat \theta) (\hat \beta_{lm} - \beta_{0}) \to_d N(0, I_p)
    \]
\end{theorem}

\subsection{Network Model Estimation} \label{sec: Network Model Estimation}

We next discuss the estimation of the generative model for the network using a variety of data types. We first demonstrate results for estimating parameters in a stochastic blockmodel. 
 We then extend the result to view the blockmodel as an approximation of a graphon. 
 \citet{breza2017using} and \citet{ breza2023consistently} consistently estimate a generative model for ARD with mutually exclusive traits. We extend this work by introducing a novel method for estimating the stochastic blockmodel with non-mutually exclusive traits using constrained least squares approach.  This innovation has major implications for practice since it can dramatically reduce survey length (since asking about multiple categories requires constructing separate questions for each intersection to make the traits mututally exclusive).  Our approach applies to a range of partial network data, not just ARD, but we summarize the resulting rates using ARD for a variety of model classes in the appendix in Table~\ref{tab: ARD rate of network model table} for comparison with existing literature.

In the main text, we concentrate on estimating the stochastic blockmodel using ARD.  In the Appendix in section~\ref{sec: estimation of SBM} we estimate generative models using partial network data such as subgraph sampling and develop similar rates for the stochastic blockmodel for subgraph sampling and reference a similar result for respondent driven sampling.

\subsubsection{SBM Estimation with ARD} \label{sec: SBM Estimation with ARD}

Recall that $X^*_{it}$ represents a set of ARD response vectors. ~\citet{breza2023consistently} show that we can consistently estimate the connection probabilities between latent types, however, we present an improved version of the SBM estimator which allows for an non-mutually exclusive traits. Let $n_t$ denote the total number of individuals of trait type $t$. Let $N'_k$ denote the nodes in our sample in group $k$, and let $n_k$ denote the number of nodes in the graph in group $k$. We cluster the node memberships according to Algorithm~\ref{alg: ARD-SBM clustering}.

\begin{algorithm} 
\footnotesize
\caption{ARD SBM clustering procedure}
\begin{algorithmic}[1]
    \State Count the number of individuals with each trait $n_t$
    \State Denote the normalized ARD responses $ X^\dagger_{it} = X^*_{it}/n_t$. 
    \State Cluster the normalized ARD response vectors $\{X^\dagger_{i}\}_{i = 1}^T$ into $K$ groups using hierarchical agglomerative clustering into a set of clusters $\hat k_i \in \{1,2,\dots, K\}$
\end{algorithmic}\label{alg: ARD-SBM clustering}
\end{algorithm}

After we obtain a clustering, we can estimate the stochastic blockmodel. Let $\hat \Omega_{kt} = \hat N_{kt}/N_t$ where $N_{kt}$ are the number of traits in the estimated group $k$ and with trait $t$, and $N_t$ are the number of individuals with trait $t$, and $\Omega_{kt} = N_{kt}/N_t$, the analogous population quantity. 
We next define the probability matrix of observing a connection of group $k$ with a trait $t$. $\mathbf{\tilde P}_{kt} = \sum_{k'} \mathbf{P}_{kk'}\omega_{k't}$, where $\mathbf{\tilde P}_{kt} = P(G_{ij} = 1|k_j = k, t_i = t)$. This relationship can be expressed in a linear system $\mathbf{\tilde P} =  \Omega \mathbf{P}$ where $\Omega \in \mathbb{R}^{T \times K}$ and $\Omega_{kt} = \omega_{kt}$. If $\Omega$ is of full column rank, then a unique solution will exist as: 
\begin{equation*}
    \mathbf{\hat P}_{kk'} = \left( \hat \Omega^\intercal \hat \Omega \right)^{-1} \hat \Omega^\intercal \mathbf{\hat{\tilde{P}}} \quad \text{ where } \mathbf{\hat{\tilde{P}}}_{kt} = \frac{1}{n_k n_t} \sum_{i \in \hat N_k} X^*_{it}. 
\end{equation*}

In general, one can symmetrize $\mathbf{\hat P}_{kk'}$ after the estimate to ensure the constraints of an undirected stochastic blockmodel are satisfied. Alternatively, once can also minimize the constrained least squares objective which can be implemented using standard convex solvers such as \texttt{CVX} \citep{Fu2020CVXR:Optimization}
\al{
    \vect{\hat P} = \argmin_{ 0 \leq \vect{P} \leq 1:\vect{P} = \vect{P}^{\intercal}  }\sum_{i = 1}^n \sum_{t = 1}^T (\tilde{X}_{it} - \sum_{k'} \widehat{\Omega}_{k' t} P_{k', k_i} )^2.
}
~\citet{breza2023consistently} develop a method for consistently estimating the stochastic blockmodel. We extend their result by obtaining a rate for estimating model parameters (Lemma~\ref{lemma: ARD_clustering}) and relax the assumption that of mutually exclusive traits. We differentiate between the estimated cross-group probabilities $\mathbf{P}^{(\vect{\hat k})}$ and those under known membership $\mathbf{P}^{(\vect{k})}$.
\begin{lemma} \label{lemma: ARD_clustering}
Suppose that we use the clustering strategy outlined in section~\ref{sec: SBM Estimation with ARD} to cluster the observations based on aggregated relational data. Let $Z_{k} = (\mathbf{\tilde P}_{k1}, \dots  \mathbf{\tilde P}_{kT})$ and $\mathbf{\tilde P}_{kt} = P(G_{ij} = 1|k_i = k, t_j = t)$. Assume also that $\inf_{k, k'} \norm{Z_{k} - Z_{k}}_2 > 0$ and that $T \geq K$ where $T$ is the number of discrete traits asked about and $K$ is the true number of clusters. 

    Let $\vect{\hat k}$ denote the estimated cluster memberships and let $\mathbf{\hat P^{(\vect{\hat k})}}$ be the corresponding estimate of the cross block probabilities. Let $\Omega_{kt} = N_{k t}/ N_t$ denote the matrix which involves the fraction of the individuals in cluster $k$ who also have trait $t$, and $\hat \Omega$ the estimated counterpart based on membership clusters. Let $C_{\Omega} = \lambda_{max}((\Omega^T\Omega)^{-1})$ and $\lambda_{max}(\cdot)$ denotes the largest eigenvalue of a matrix and $C_{\Omega} > 0$. Then with probability at least $1 - \delta - \frac{1}{n}$
    $$ \norm{\mathbf{\hat P^{(\vect{\hat k})}} - \mathbf{P}^{(\vect{k})}}_1 \leq  C_{\Omega}  \frac{KT}{n} \sqrt{\frac{\log(2/\delta) \log(KT)}{2}}$$
\end{lemma}

We contrast our results to the optimal estimation rate for a stochastic blockmodel from \citet{gao2015rate}, $\tilde O_P(n^{-1/2})$. Our rate appears faster due to the complexity difference in clustering problems. Our clustering benefits from node-level traits, which provide extra information. As the network grows, the normalized ARD vector converges to its mean, simplifying clustering and resulting in a faster rate.

\subsubsection{Misspecification of the Graph Model} \label{sec: Graphon}

We use a stochastic blockmodel as it effectively approximates a general graphon class. Even if $\theta_0$ belongs to a smooth graphon class rather than a stochastic blockmodel, we can still bound the bias in estimating the relevant model parameters. 
Consider a scenario where edges are generated under a true graphon model $\tilde g$ where $\eta_{ij} =  \tilde g(\xi_i, \xi_j) = P(G_{ij} = 1|\vect{\xi}) \text{ where } \vect{\xi} \sim_{iid} P_{\vect{\xi}} \in [0,1]. $
Let $\mathcal{H}_{\alpha}(M)$ denote a smooth graphon class defined via the $\alpha$-$M$-H\"{o}lder class as follows. Let $\mathcal{D} = [0,1]^2 \cap x \leq y$ denote the domain of $(x,y)$. We define the norm $\norm{\tilde g}_{\mathcal{H}_\alpha}$ as: 
$$ \norm{\tilde g}_{\mathcal{H}_\alpha} = \max_{j + k \leq \lfloor \alpha \rfloor} \sup_{x,y \in \mathcal{D}} |\nabla_{jk} \tilde g(x,y)| + \max_{j + k = \lfloor \alpha \rfloor}\sup_{(x,y) \not = (x'y') \in \mathcal{D}}\frac{\nabla_{jk} \tilde g(x,y) - \nabla_{jk} \tilde g(x',y')}{(|x- x'| + |y - y'|)^{\alpha - \lfloor \alpha \rfloor}}$$
and the H\"{o}lder class corresponding to this norm as
$$ \mathcal{H}_\alpha(M) = \{\norm{\tilde g}_{\mathcal{H}_\alpha} \leq M: \tilde g(x,y) = \tilde g(y,x); 0 \leq \tilde g(x,y) \leq 1\}. $$
Prior work has focused on the approximability of a stochastic blockmodel to any element of a smooth graphon class. In particular there will always be some assignment of block memberships such that we can bound the $2$-norm probability deviation from the true model. 

\begin{lemma} \label{lemma: graph misspecification}
    Suppose that $\theta_*$ corresponds to a true graphon model and $\theta_0$ a corresponding approximating stochastic blockmodel. Denote the population estimating function, as a function of the model parameters
    \al{
        L_n(\beta, \theta) &= \E[\tilde m_{n}(\vect{\tilde Z}; \beta) |\vect{a}, \vect{X}, \theta]
    }
    where $L_n(\beta_0, \eta_0) = 0$ defines the population parameter $\beta_0$ under the misspecified model $\theta_0$, and let $L_n(\beta_*, \theta_*) = 0$ define the population solution $\beta_*$ to the correctly specified graph model $\theta_*$. Let $\eta_{0}$ and $\eta_*$ be the pairwise edge probabilities corresponding to the models $\theta_0, \theta_*$ respectively. 
    Finally assume that: 
    \begin{enumerate}[label=D\arabic*.,ref=D\arabic*]
        \item $\mathcal{B}$ is compact
        \item $\sup_{\beta \in {\mathcal{B}}} |L_n(\beta, \eta) - L_n(\beta, \eta_{*})| \leq L\norm{ \eta -  \eta_*}_2/n$ 
        \item $\min_{j} \frac{\partial}{\partial \beta_j}L_n(\beta,\eta_*) \bigg |_{\beta = \beta_*} = \lambda > 0$
    \end{enumerate}
    Then the approximation error under the graph misspecification is bounded by the rate: 
    \begin{equation}
        \norm{\beta_0 - \beta_*} = O(\lambda^{-1}K^{-\alpha\wedge 1}) \quad \text{where $a\wedge b = \text{min}(a,b)$.}
    \end{equation}
\end{lemma}

In practice, we don't directly select clusters; misspecified clusters relate to observed traits, thus this bound holds only under good alignment of clusters. This bound is a worst-case scenario and may be overly conservative regarding observed bias. Future work could involve sensitivity analysis of the response function and the latent graph model.

\section{Experimental Design} \label{sec: experimental design}

So far, our focus has been on estimating model parameters given a treatment assignment $\vect{a}$. We now explore experimental design methods that leverage partial network data to choose $\vect{a}$ to minimize the variance of our estimands.  Leveraging partial network data for this purpose is particularly appealing in practice, since it requires substantially less investment than collecting full network data and could be collected as part of creating a sampling frame in settings where researchers collect data to construct the frame.

We consider saturation randomization experiments, which divide the dataset into $J$ clusters of size $n_j$. A proportion $\tau_j$ of each cluster is assigned the treatment, totaling $n_t = \sum_{j = 1}^{J} \tau_j n_j$, and generally will not be the same ``blocks" as those in a graph model if the graph model uses discrete factors (e.g. stochastic blockmodel).  Practically, due to budget constraints, the set of possible saturation levels $\vect{\tau}$ is limited to $\mathcal{T} \subset [0,1]^{J}$. For example, this could be due to limited resources like a finite vouchers in a vaccine trial.

\subsection{Bayesian Optimization of Asymptotic Regression Estimators}
Our goal is to optimize the asymptotic variance of a function of the model parameter $\hat \beta$ in section~\ref{sec: Inference}. We highlight this by optimizing the variance of the estimates of linear contrasts of the parameters $\phi^T \beta$. When using the stochastic block model for the network model these treatment blocks could align with the model blocks, however this need not (and likely won't) be the case.  They could, instead, be based on observed characteristics (e.g. geography, classrooms). 

Denote the variance of the target contrast parameter conditional on the treatment assignment as $\vect{a}$: $\upsilon^{\phi}(\vect{a}; \theta) = \Var(\phi^T \hat \beta|\vect{a}, \theta)$. Ideally, the goal is to find a treatment assignment $a^*$ that minimizes the variance of the contrast: $a^* = \argmin_{\vect{a} \in \{0,1\}^n} \upsilon^{\phi}(\vect{a}; \theta).$
Without added structure, optimizing treatment assignments is NP-hard, requiring a search over $2^n$ possible assignments. By changing the objective to one where we optimize over a set of saturation levels over a set of groups $\vect{\tau} \in [0, 1]^{J}$, we simplify the problem so that it is no longer NP-hard (i.e. since $J \ll n$ typically) and is therefore tractable. The distribution of treatment assignments, $\vect{a}$, under $\vect{\tau}$ is denoted by $P_{\vect{\tau}}$, and we aim to minimize:
$$\V(\vect{\tau}; \theta) = \E_{\vect{a} \sim P_{\vect{\tau}}}[\upsilon^{\phi}(\vect{a}; \theta_0)].$$

In Algorithm~\ref{alg: Saturated Random Design Variance OLS}, we present a method for evaluating the variance of a linear model using a generic feature map $\tilde h$ for a given treatment assignment $\vect{a}$ and a graph model $\theta$. A general approach for Z-estimators is detailed in the Appendix. Algorithm~\ref{alg: Saturated Random Design Variance OLS} operates under specific assumptions about the covariance matrix $\Sigma$, which may include correlations within densely connected network components. We will present our algorithm for minimizing this variance using Bayesian optimization, which accounts for the uncertainty in the outcome, given a graph model. In the appendix we give an extension which also  which incorporates network model uncertainty $\hat \theta$ (section~\ref{sec: Variance Minimization with Model Uncertainty}).

\begin{algorithm} 
\footnotesize
\caption{Saturation Randomized Design Variance.}
\begin{algorithmic}[1]
    \State \textbf{Inputs: } Variance structure $\Var[\vect{u}] = \Sigma$, Model estimate $\hat \theta$. 
    \State Sample $L$ draws from the graph model $\{\hat{G}^{(l)}\}_{l = 1}^L \sim \hat \theta|G^*$ 
    \State Sample $R$ treatments $\{\vect{a}_{r}\}_{r = 1}^R$ according to the block saturation levels $\vect{\tau}$. 
    \For{$r \gets 1$ \textbf{to} $R$} 
    \State Compute the averaged features over draws from the graph model $\{\hat{G}^{(l)}\}_{l = 1}^L$, 
    $$ \hat H_{ir}(\vect{a}) = \frac{1}{L} \sum_{l = 1}^L \tilde h(S_i(\hat{G}^{(l)}) V_i(\vect{a}_r; \hat{G}^{(l)})) $$
    \State Compute the Hessian $ \mathsf{\hat H}_n(\vect{a}_r) = \frac{1}{n} \sum_{i = 1}^n \hat H_{ir}(\vect{a}) \hat H^T_{ir}(\vect{a})$. 
    \State Compute the design matrix $\hat H^T_{r}(\vect{a}) \in \R^{n \times p}$ where each row is $\hat H_{ir}(\vect{a})$. 
    \State Compute the variance for a single draw of the treatment vector $\vect{a}_r$: 
    $$\upsilon^{\phi}(\vect{a}_r; \hat \theta) = \phi^T \mathsf{\hat H}^{-1}_n(\vect{a}_r)   \hat H^T_{r}(\vect{a})\Sigma \hat H_{r}(\vect{a})\mathsf{\hat H}^{-1}_n(\vect{a}_r) \phi $$
    \EndFor
    \State Average over each of the draws $\V(\vect{\tau}; \hat \theta) = \sum_{r = 1}^R \upsilon^{\phi}(\vect{a}_r; \hat \theta)$
\end{algorithmic}\label{alg: Saturated Random Design Variance OLS}
\end{algorithm}

\textbf{Bayesian Optimization}. 
Calculating the average variance $\V(\vect{\tau}; \hat \theta)$ in Algorithm~\ref{alg: Saturated Random Design Variance OLS} is computationally intensive to evaluate and often non-convex. Since the number of cluster saturation tends to be relatively small, this suggests that Bayesian optimization is an appropriate method for minimizing this saturation variance. 
Let $\V(\tau) := \V(\tau; \hat \theta)$ denote our objective function of the variance evaluated using an estimate of the network model $\hat \theta$. Given a set of pilot points $\vect{\tau}_1, \vect{\tau}_2, \dots, \vect{\tau}_{n_0}$ (i.e. uniformly sampled on $\mathcal{T}$) we propose a Gaussian process prior satisfying 
\begin{equation*}
    \V(\vect{\tau}_{1:n_0}) \sim N(\mu_0(\vect{\tau}_{1:n_0}, \Sigma_0(\vect{\tau}_{1:n_0}, \vect{\tau}_{1:n_0})))    
\end{equation*}
where $\Cov[\V(\vect{\tau}_{i}), \V(\vect{\tau}_j)] = \Sigma_0(\vect{\tau}_i,\vect{\tau}_j)$ where $\Sigma_0$ is a positive semidefinite kernel function. As a default, we use the Gaussian kernel
$\Sigma_0(x,x') = \alpha_0 \exp\left( -\norm{x - x'}^2\right)$. We can then use this prior to define a posterior over remainder of the design space $\mathcal{T}$
\begin{align*}
    \V(\vect{\tau})|\V(\vect{\tau}_{1:n_0}) &\sim N(\mu_n(\vect{\tau}), \sigma^2_n(\vect{\tau})) \\
    \mu_n(\vect{\tau}) &= \Sigma_0(\vect{\tau}, \vect{\tau}_{1:n_0}) \Sigma_0(\vect{\tau}_{1:n_0}, \vect{\tau}_{1:n_0})^{-1}(\V(\vect{\tau}) - \mu_0(\vect{\tau}_{1:n_0})) + \mu_0(\vect{\tau}) \\
    \sigma^2_n(\vect{\tau}) &= \Sigma_0(\vect{\tau}, \vect{\tau}) - \Sigma_0(\vect{\tau}, \vect{\tau}_{1:n_0}) \Sigma_0(\vect{\tau}_{1:n_0}, \vect{\tau}_{1:n_0})^{-1} \Sigma_0(\vect{\tau}_{1:n_0}, \vect{\tau}).
\end{align*}

From this posterior, we define an acquisition function $A(\vect{\tau})$.  As a default, we choose the upper confidence bound (UCB) acquisition function  $A(\vect{\tau}) = \mu_n(\vect{\tau}) - \kappa \sigma_n(\vect{\tau})$ for a chosen $\kappa$ (where we set $\kappa = 2$). This method is implemented in the \texttt{R} package \texttt{rBayesianOptimization}, which uses \texttt{GPfit} \citep{rBayesianOptimization_package, macdonald2015gpfit}. For a detailed review of Bayesian optimization techniques, refer to \citet{frazier2018tutorial}.  We evaluate the complete Bayesian optimization procedure in Algorithm~\ref{alg: Bayesian Optimization}, where we apply the procedure for $N_0$ iterations.

\begin{algorithm} 
\footnotesize
\caption{Bayesian Optimization Procedure}
\begin{algorithmic}[1]
    \State \textbf{Inputs: } Graph model $\hat \theta$ and partial graph information $G^*$. Kernel function $\Sigma_0$. 
    \State Sample $\vect{\tau}_{1:n_0}$ uniformly from $\mathcal{T}$, as a pilot sample of the design points. 
    \State Update the posterior on $\V(\vect{\tau})$. 
    \For{$i \gets n_0 + 1$ \textbf{to} $n_0 + N_0$} 
    \State Update the posterior on $\V(\vect{\tau})|\V(\vect{\tau}_{1:(i - 1)})$.
    \State Let $\vect{\tau}_i$ be the minimizer of the acquisition function $A(\vect{\tau})$ (UCB). 
    \State Evaluate $\V(\vect{\tau}_i)$ using Algorithm~\ref{alg: Saturated Random Design Variance OLS}. 
    \EndFor
    \State Return the point $\vect{\tau}_{1:(n_0 + N_0)}$ with the smallest $\V(\vect{\tau})$
\end{algorithmic}\label{alg: Bayesian Optimization}
\end{algorithm}

The quality of optimization over $N_0$ iterations depends on the smoothness of $\V(\vect{\tau})$. Since variance might diverge under some settings (e.g., as $\vect{\tau} \to 0$), a simple alternative is to maximize $\exp(-\V(\vect{\tau}))$ instead. The closeness of the maximizer after $N_0$ iterations hinges on the smoothness of $\exp(-\V(\vect{\tau}))$, which we assume belongs to a reproducing kernel Hilbert space, $\mathcal{H}$, with a bounded kernel $\Sigma_0(x,x') \leq B$. This function's smoothness affects the approximation rate, detailed in \citet{srinivas2009gaussian}. For instance, with Gaussian kernel $\Sigma_0$, the approximation error is $\exp(-\V(\vect{\tau}^*)) \geq \frac{1}{N_0}\sum_{m = 1}^{N_0}\exp(-\V(\vect{\tau}_m)) + O_P(\frac{B\sqrt{\log(N_0)^{K + 1}} + \log(N_0)^{K + 1}}{\sqrt{N_0}})$. Similar findings apply to Matern and linear kernels per \citet{srinivas2009gaussian}.

\subsection{Designs for Optimal Seeding}
Given a model of the potential outcomes, we may also leverage this model for optimal seeding, a task that is NP-hard \citep{kempe2003maximizing} in general. Many contagion models are exchangeable given an exposure, and with only block information available, then we can reduce our search space to that over block saturation. In our case, where exact network structures are unknown, we determine the optimal blocks for seeding. When $K \ll n$, this structure significantly reduces computational efforts, and we only need to decide how many seeds to allocate to each of the $K$ clusters.

The model leveraged for the outcome $f_Y(V_i, S_i, \vect{\varepsilon}_Y)$ could be a predefined model based on domain knowledge, such as complex contagion used by \citet{beaman2021can}. In other scenarios, this might be estimated (e.g., simulation using $f_Y(V_i, S_i, \vect{\varepsilon}_Y; \widehat \beta)$ in place of $f_Y(V_i, S_i, \vect{\varepsilon}_Y)$). This is demonstrated in Algorithm~\ref{alg: Optimal Seeding Partial Network} (line 5).

\begin{algorithm} 
\footnotesize
\caption{Optimal Seeding With Partial Network Data}
\begin{algorithmic}[1]
    \State \textbf{Inputs: } Number of seeds $N$, Model estimate $\hat \theta$, number of graph draws $L$. 
    \State Sample $L$ draws from the graph model $\{\hat{G}^{(l)}\}_{l = 1}^L \sim \hat \theta|G^*$ 
    \For{ $\tau \in \mathcal{T}$} 
    \State Sample $L$ treatments $\{\vect{a}_{l}\}_{l = 1}^L$ according to the block saturation levels $\vect{\tau}$. 
    \State Compute the outcomes $Y^{(l, \vect{a}_{l})}_i$ according to the outcome model $f_Y(V_i, S_i, \vect{\varepsilon}_Y)$. 
    \State Compute the average (and standard error) over draws of the network $\bar{Y}^{(\vect{\tau})} = \frac{1}{L}\sum_{l = 1}^L Y^{(l, \vect{a}_{l})}_i$
    \EndFor
    \State Return saturation level $\vect{\tau}$ with the largest value of $\bar{Y}^{(\vect{\tau})}$. 
\end{algorithmic}
\label{alg: Optimal Seeding Partial Network}
\end{algorithm} 
When the total number of seeds (see Algorithm~\ref{alg: Optimal Seeding Partial Network}, line 3) is small, it is computationally feasible to implement exactly. Alternatively, we could use Bayesian optimization to control treatment saturation levels.

\section{Data Analysis} \label{sec: Data Analysis}

In this section, we present three empirical examples to illustrate our framework's utility in estimating causal effects, designing experiments, and implementing seeding strategies. We adopt a semi-synthetic approach in our examples, where the outcomes are simulated based on processes derived from real networks. The networks analyzed pertain to observational and experimental studies focused on information diffusion in rural villages in India and Malawi, as discussed in \citet{banerjee2013diffusion}, \citet{banerjee2019gossip}, and \citet{beaman2021can}. These networks consist of 30-400 households per village.  To ensure continuity across the examples, we generate ARD as the partial data type and model the networks using stochastic blockmodels for each case, however the use of other network generative models and partial network datatype are applicable in these cases. 

When covariates are available for all nodes, we use them to construct ARD. If covariates are missing, we apply the Leiden algorithm \citep{traag2019louvain} in \texttt{igraph} \citep{csardi2006igraph} to cluster the network and treat these clusters as traits. Table~\ref{tab: ARD Examples} details which datasets used actual traits versus clustering to manage trait numbers in our simulations.
\begin{table}[htbp] 
\footnotesize
\centering
\begin{tabular}{|l|c|}  
\hline 
Network Dataset & Traits               \\ \hline
\citet{banerjee2013diffusion}  & Leiden Cluster $K \in [4,16]$\\
\citet{banerjee2019gossip}   & Observed Traits (section~\ref{sec: Gossips ARD Questions})          \\
\citet{beaman2021can}     & Leiden Cluster K = 8 \\
\hline 
\end{tabular}
\caption{Summary of synthetic traits vs. real traits in the semi-synthetic simulations.}
\label{tab: ARD Examples}
\end{table} 
In Section~\ref{sec: causal effect estimation}, we use networks from \citet{banerjee2013diffusion}, which include various social relations from 70 villages, each with 80 to 350 individuals per household. In Section~\ref{sec: example Experimental Design Simulations}, we use networks from \citet{banerjee2019gossip}, consisting of 68 similar-sized villages, and repeat the simulations 500 times per village. In Section~\ref{sec: Optimal Seeding}, we include networks from \citet{beaman2021can}, excluding those with insufficient connections for diffusion. This leaves 114 villages with 30 to 350 households per village, and we repeat the simulations 2000 times per village.

\subsection{Causal Effect Estimation} \label{sec: causal effect estimation}

In our first example, the aim is to estimate the global average treatment. We consider the example from \citet{ugander2023randomized} and generate a set of potential outcomes according to the following model 
$$ Y_i(\vect{0}) = \frac{d_i}{\bar d}\cdot \left(\alpha + bX_i + \sigma \epsilon_i \right), \quad Y_i(\vect{a}) = Y_i(\vect{0})\cdot\left(1 + \delta a_i + \gamma\frac{\sum_{j \in [n]} G_{ij}a_j}{d_i} \right) $$
where $\epsilon_i \sim_{iid} N(0,1)$ is some independent noise, and $X_i$ is a covariate that varies throughout the network, $d_i$ is the degree of individual $i$ and $\bar d$ is the average degree across the network. We set $\alpha = 1$, $b = 1$, $\delta = 1$, $\sigma = 0.5$ and $\gamma = -0.5$. The global average treatment effect in this model is $\frac{1}{n}\sum_{i = 1}^nY_i(\vect{0})(\delta + \gamma) = \Psi(\vect{a} = 1|G) - \Psi(\vect{a} = 0|G)$. The exposure is the individual treatment in conjunction with the average treatment of neighbors, and the graph confounder include the degree ratio and node level covariates
$$f_V(\vect{a}; \varphi_i(G)) = \left(a_i,  \frac{\sum_{j \in [n]} G_{ij}a_j}{\bar d}\right), \quad f_{S}(\vect{X};\vartheta_i(G)) = \left(\frac{d_i}{\bar d}, X_i\right). $$  

\begin{figure}[H]
\centering
\subfigure[Bias]{
\includegraphics[width = 0.45\textwidth]{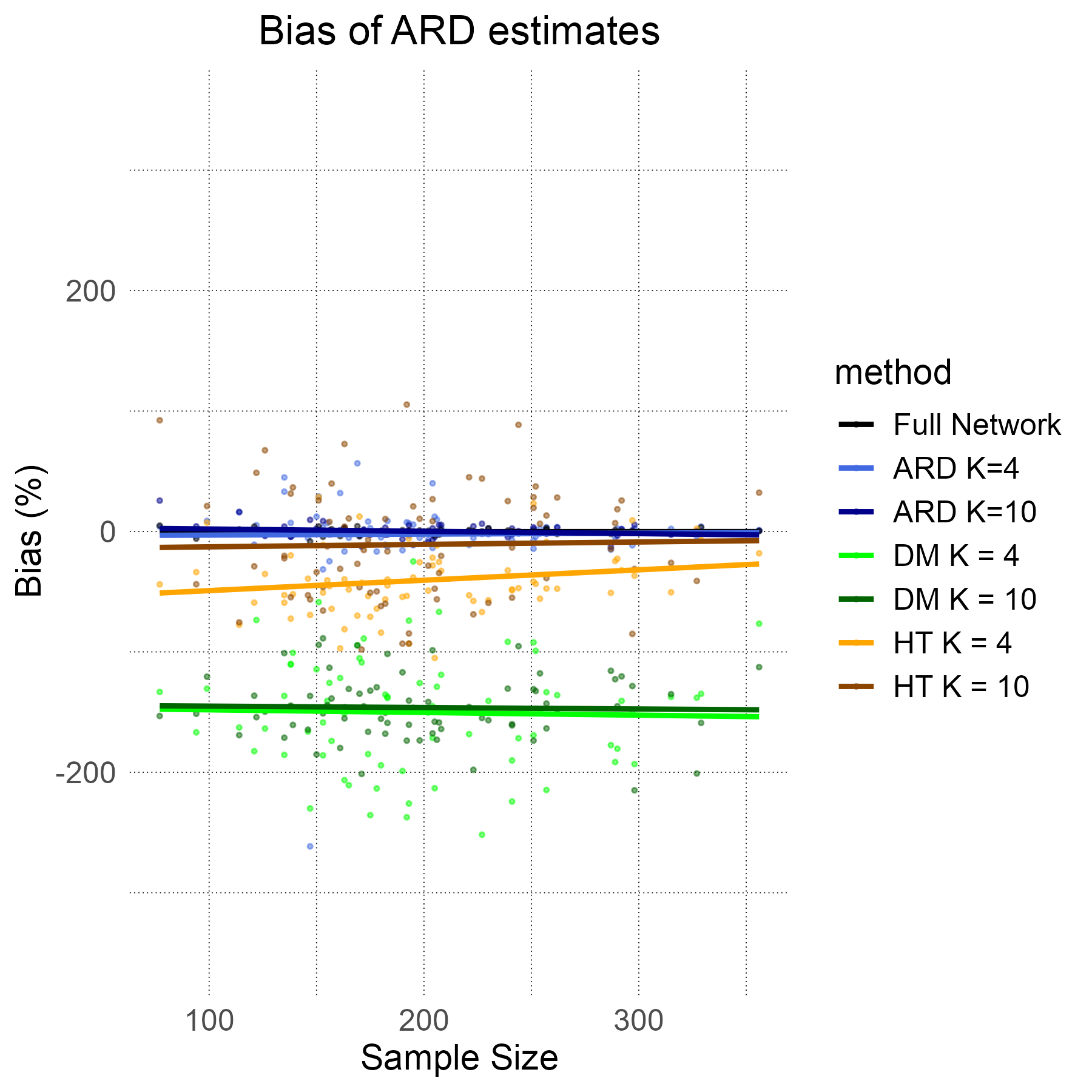}
\label{fig: Ugander GATE bias}
}
\subfigure[RMSE]{
\includegraphics[width = 0.45\textwidth]{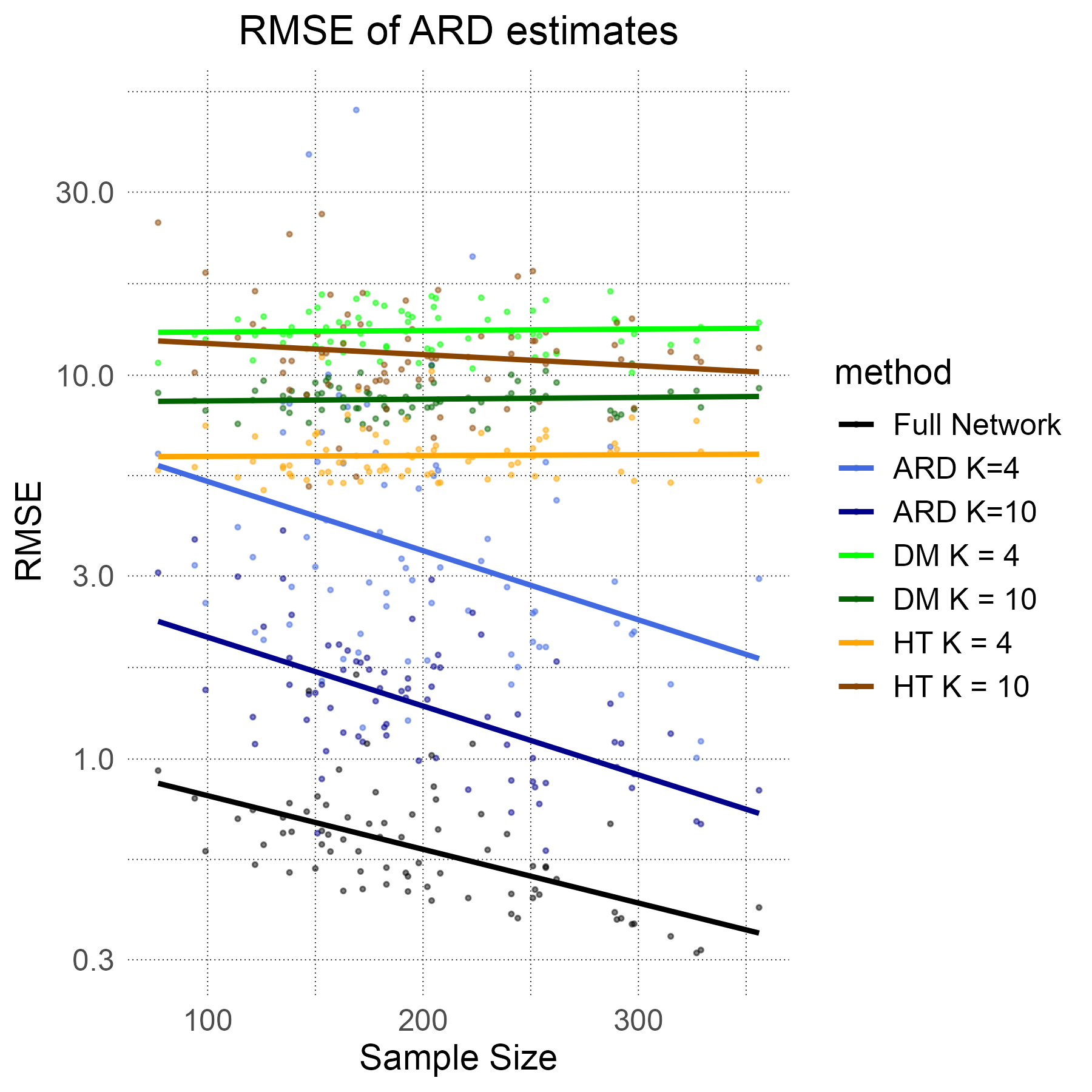}
\label{fig: Ugander GATE rmse}
}
\caption{Comparison of GATE estimators. ARD denotes our method using aggregated relational data. The ``Full Network" method uses a regression approach with the full data available. DM is the difference in means and HT is the Horvitz-Thompson estimator.}
\label{fig: Ugander GATE}
\end{figure}

We evaluate the effectiveness of graph cluster randomization by comparing a Horvitz-Thompson estimator \citet{Ugander2013GraphUniverses} to a difference in means estimator under a cluster randomized design. In this design, half of the clusters receive no treatment (saturation of $0$) and the other half receive full treatment (saturation of $1$). We vary the number of clusters from 4 to 16 but display results only for 4 and 10 clusters in Figure~\ref{fig: Ugander GATE} for clarity.

Figure~\ref{fig: Ugander GATE} shows that the full data regression model performs the best, as it leverages more information than the ARD approaches. However, the ARD version still effectively minimizes bias (Figure~\ref{fig: Ugander GATE bias}) and RMSE (Figure~\ref{fig: Ugander GATE rmse}). In our simulations of dense graphs with few clusters, the Horvitz-Thompson Estimator faces challenges as the network grows—almost all nodes have at least one neighbor with a treatment different than their own. The difference in means estimator shows consistent bias, due to not using heterogeneous covariate information. While regression with complete data is most effective, using partial network data still yields comparably good results.

\subsection{Experimental Design} \label{sec: example Experimental Design Simulations}

We next highlight aspects of experimental design using an information diffusion example based on the hearing model referenced in section~\ref{sec: examples of exposure maps}. At each time step the previously infected nodes are susceptible again the nodes infected in the last round will infect their neighbors with probability $q_{t + 1}$. We repeat this for $T = 3$ rounds. Let $N_i$ denote the total number of infections after the process. We then sample some binary response $P(Y_i = 1|N_i) = \text{logit}(\alpha_0 + \alpha_1 N_i)$ where $\alpha_0$ and $\alpha_1$. 

In this case, $V_i = \E[N_i|\vect{a}] = \sum_{t = 0}^3 \beta_t\vect{a}(G^{t})_{i}$ where $\beta_t = \prod_{j = 1}^t q_j$. We estimate the coefficients in each of these cases letting $V_i = \E[N_i|\vect{a}]$ be the exposure mapping. We then generate the outcomes according to the exposure received 
\al{
    \E[Y_i|S_i,V_i] = \Lambda(\alpha_0 +\alpha_1 (\sum_{t = 0}^3 \beta_t(G^{t})_{i} \vect{a}))
}
where $\Lambda(\cdot)$ is the logistic function. For our experiments, we set $\beta = (0,0.5,0.05,0.005)$.

\begin{figure}[H]
\centering
\subfigure[Estimation of $\alpha_1$.]{
\includegraphics[width = 0.45\textwidth]{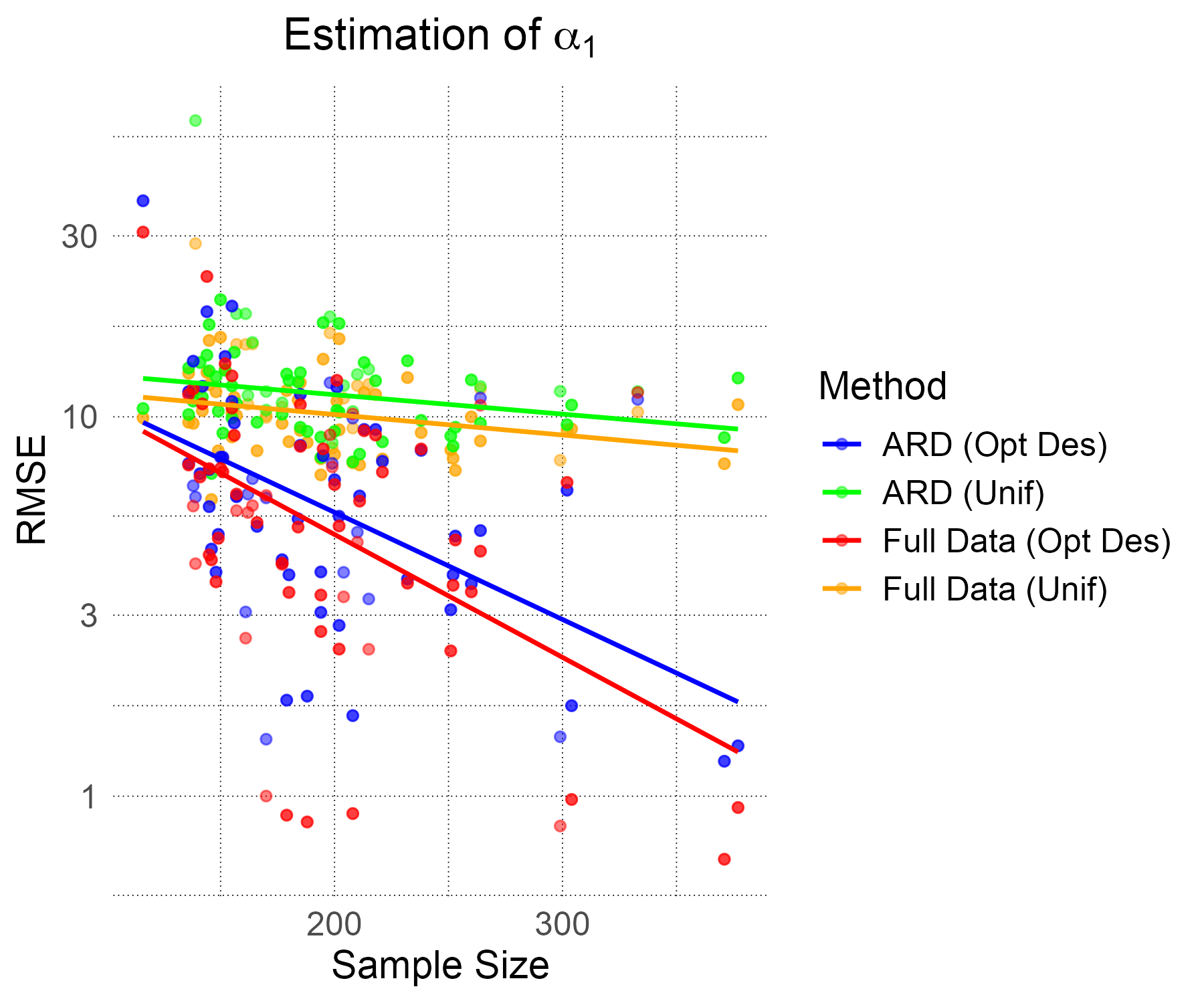}
\label{fig: Estimation of alpha1}
}
\subfigure[Estimation of all model parameters.]{
\includegraphics[width = 0.45\textwidth]{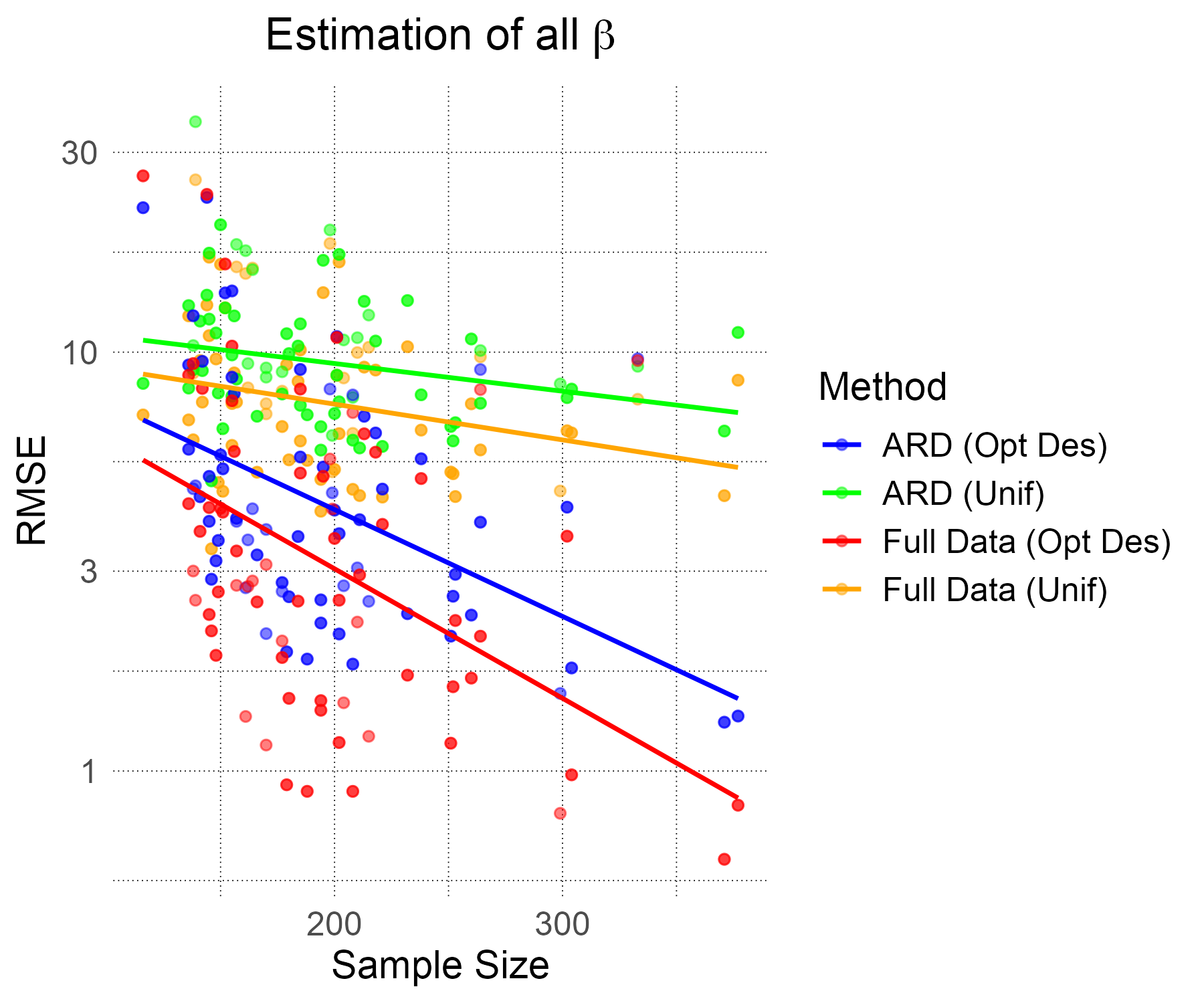}
\label{fig: Estimation of beta}
}
\caption{Estimation of parameter $\alpha_1$ and all model parameters $\beta$ using the naive and optimized seeding. We observe that the potential gain found using a more efficient design is much greater than simply collecting complete network data.}
\label{fig: Gossip Hearing Model}
\end{figure}

In the dataset, seeds are assigned uniformly with either 3 or 5 seeds per network. Following our procedure in section~\ref{sec: experimental design}, we compute the optimal seed allocations, ensuring no cluster receives more seeds than available in the actual experiment (either 3 or 5). In practice our Bayesian optimization procedure starts by randomly sampling the target space 20 times, followed by 20 iterations to refine saturation. We then compare the estimates for $\alpha_1$ and all model parameters as shown in Figure~\ref{fig: Gossip Hearing Model}. The results indicate that a more strategically designed experiment generally yields more significant gains than directly using the graph parameters.  On average, using optimized designs rather than uniform random designs when collecting network data significantly reduced RMSE. Specifically, for estimating $\alpha_1$, the optimized design decreased RMSE by 38\% ($\pm$12\%) compared to 11\% ($\pm$2\%) with complete data (where the brackets refer to the 95\% confidence interval of the mean estimate across simulations). For all parameters, the optimized design resulted in a 45\% ($\pm$10\%) reduction in RMSE, versus an 18\% ($\pm$2\%) reduction with complete data.

\subsection{Optimal Seeding} \label{sec: Optimal Seeding}

We apply our methodology to the seeding problem described in \citet{beaman2021can}, where the diffusion of pit-planting technology among Malawian farmers follows a complex contagion process. The outcome model is defined as $Y_i = f_Y(S_i, V_i, \vect{\epsilon}_Y)$, with individuals having a threshold $\varsigma_i \sim N_{[0,\infty)}(\lambda, 0.1)$ for spreading infection based on neighbor infections from the previous time (where $N_{(a,b)}(\mu,\sigma)$ refers to the $\mu,\sigma$ normal distribution truncated on the interval $(a,b)$) . This process is simulated over three time periods to align with their experimental design, setting $\lambda = 2$ and repeating 2000 times for $K = 8$ clusters to determine optimal seeding groups.

\begin{figure}[htbp]
\centering
\subfigure[Adoption ratio as a function of edge density.]{
\includegraphics[width = 0.45\textwidth]{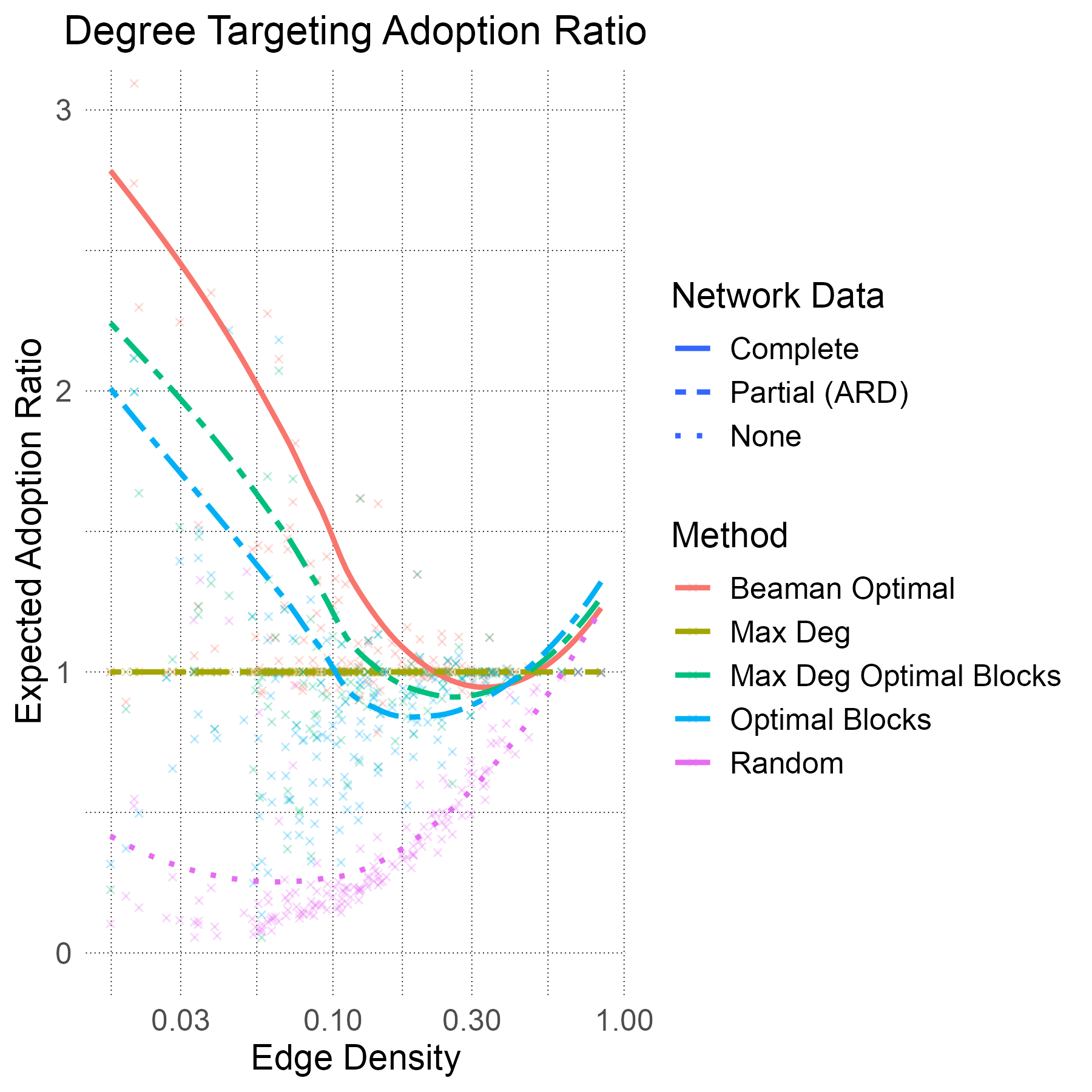}

}
\subfigure[Adoption ratio as a function of village size.]{
\includegraphics[width = 0.45\textwidth]{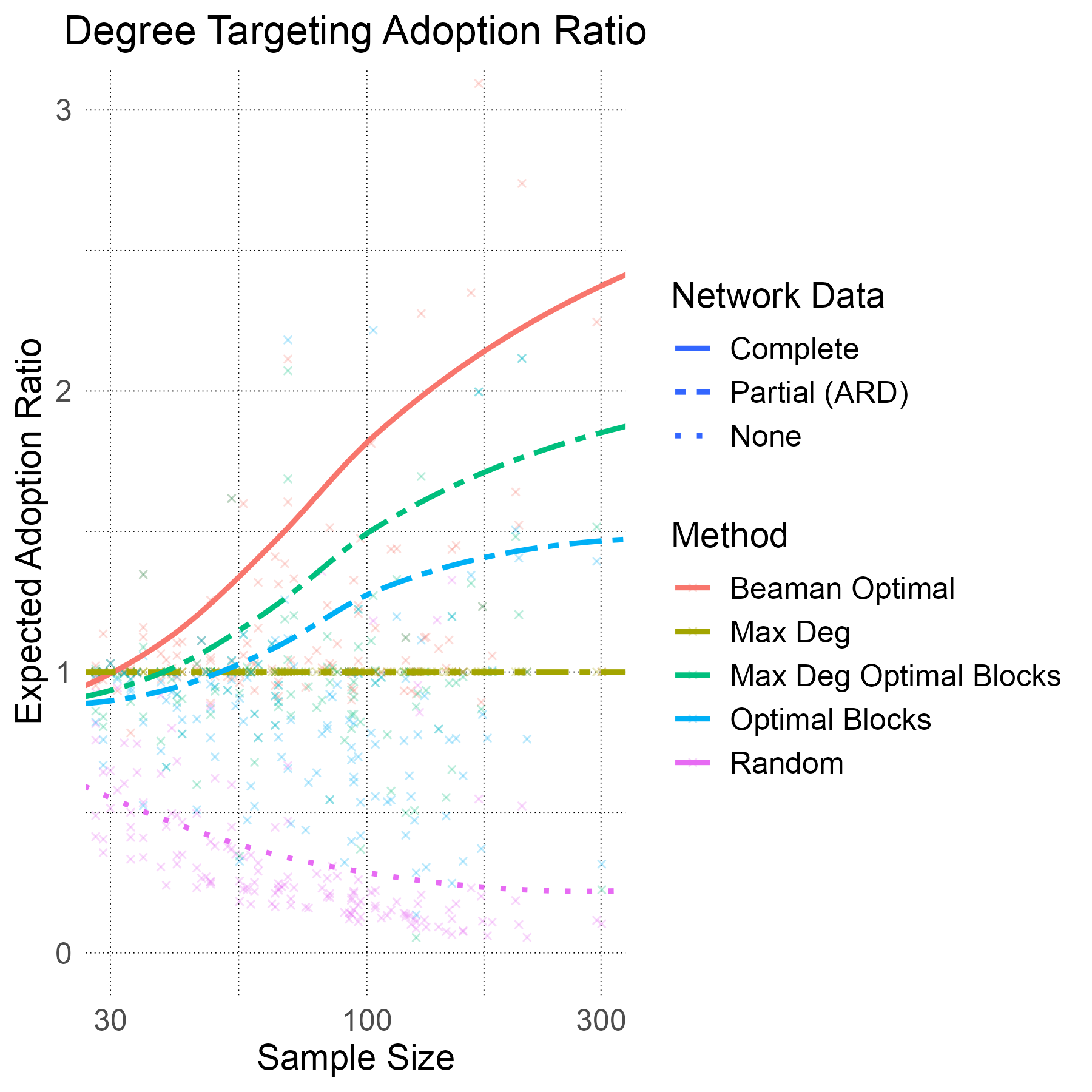}

}
\caption{Comparison of different seeding methods under complex contagion. Model-based targeting of optimal blocks generally outperforms degree seeding, especially when targeting the highest degree nodes within those blocks.}
\label{fig: complex contagion seeds}
\end{figure}

We explore two seeding strategies: randomly assigning seeds to the top two members of optimal clusters, and seeding the nodes with the highest degrees within these clusters. We compare these strategies to common degree targeting, noting that our max degree method typically yields the highest adoption rates, especially in larger, sparser villages, as illustrated in Figure~\ref{fig: complex contagion seeds}. However, in very small or dense networks, the performance differences between strategies are negligible. Across all graphs we find the optimal seeding strategy to increase adoption by $1.50$ $(\pm 0.16)$ times relative to degree seeding, while the optimal blocks was $1.13$ $(\pm 0.12)$ times and optimal degree within blocks increased adoption by $1.28$ $\pm (0.13)$ times. 

\section{Conclusions}
\label{sec:conclusions}

We introduce a framework that identifies causal effects under interference using a structural causal model, facilitating inference with partial network data. The framework is general and can be applied using broad class of outcome models and graph models. Our outcome modelling approach leveraging node-level heterogeneity and exposure mappings allow for the estimation of all causal effects, rather that other methods which tend to focus on a single causal effect like the GATE. Demonstrations through semi-synthetic problems highlight its effectiveness, matching or surpassing fully observed data methods in certain scenarios.

Our method highlights that directly modeling interference mechanisms offers several advantages, including leveraging transportability of outcome models for seeding and inference for experimental designs when estimating effects under interference.

Future studies might consider semiparametric approaches to estimation with partial data like those in \citet{auerbach2022identification}. Additional structured assumptions on potential outcomes as suggested in \citet{belloni2022neighborhood} could also be explored. Currently, our focus has been on analyzing problems at a single time point. However, future research could extend to designing experiments with panel data and staggered rollouts. It would also be worthwhile to develop classes of outcome models that more explicitly incorporate this temporal structure. 

\bibliographystyle{chicago} 

\bibliography{references}

\newpage
\bigskip
\begin{center}
{\large\bf SUPPLEMENTARY MATERIAL}
\end{center}
\renewcommand{\thesection}{A.\arabic{section}}
\numberwithin{equation}{section}
\setcounter{section}{0}
%\section{Additional Examples of Exposure Maps}

\section{Comparing Frameworks of Interference} \label{sec: causal frameworks}

We contrast the approaches of a fixed outcome approach as in \citet{Aronow2017EstimatingExperiment} to a structural causal model approach. In the former approach, each individual has a distinct outcome under an exposure $v$, $Y_i(v)$. Though such an approach is robust for learning parameters such as average treatment effects $\frac{1}{n} \sum_{i = 1}^n Y_i(v)$, the information in an individual $i$'s potential outcome is completely distinct from individual $j$. This important details has important downstream implications.

\begin{figure}[H]
\centering

\includegraphics[width = 0.6\textwidth,trim={1cm 0cm 0cm 0cm},clip]{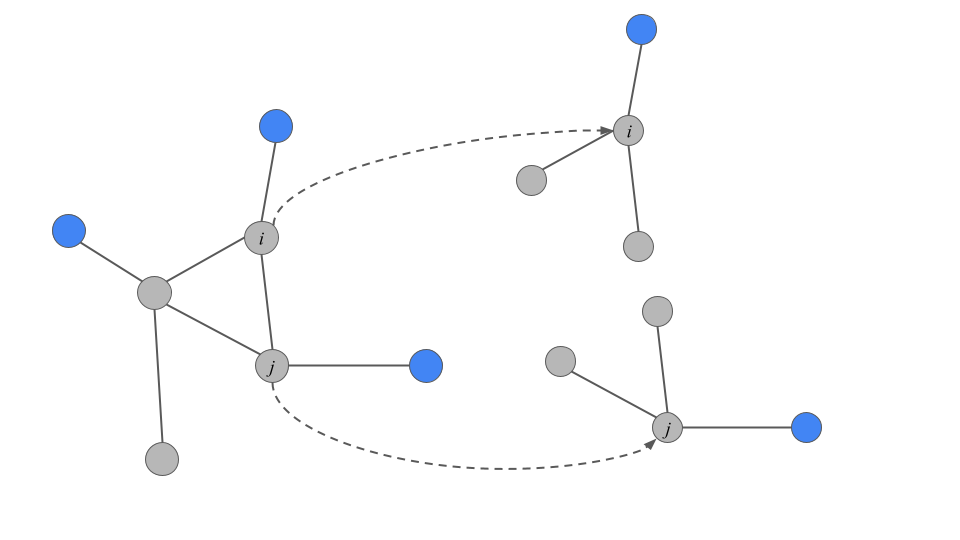}

\caption{Equivalence of distribution of potential outcomes of nodes $i$ and $j$ are equivalent under this given treatment assignment as all of the rooted networks are equivalent.}
\label{fig: Simple Contagion Rooted Network}
\end{figure}

Consider the simple contagion model from the example in section~\ref{sec: example contagion} which takes place in a single time period ($T = 1$). Consider the nodes $i,j$ in Figure~\ref{fig: Simple Contagion Rooted Network} with seeded nodes in blue. Suppose that at time $T = 1$, that each neighbour of a treated node is infected with probability $q$. Since each one has only a single treated neighbor the distribution of the infection probability $P(Y_i = 1|\vect{a}, G)$ $i$ and $j$ are equivalent as their exposures are identical (i.e. they are each connected to a single seed node). However, in the finite sample framework the potential outcomes of any two nodes with a single treated neighbor can be arbitrarily different ($Y_i(v) \not = Y_j(v)$). 

This nonparametric structure imposed on the potential outcomes later imposes restrictions on the degree of influence of others a node can have for estimation, thereby limiting this framework to examples with local dependencies (a phenomena also seen in \citet{ogburn2022causal}). 

\subsection{Why not IPW estimators?}

In many nonparametric approaches to estimating causal quantities under interference, inverse probability weighted (IPW) estimates can be developed given a randomization scheme, i.e. distribution of the assignments $P(\vect{a})$ \citep{Aronow2017EstimatingExperiment}. This is useful as it can be used to develop estimators for causal effects any exchangeability assumptions on the potential outcomes. However when $V_i$ is not observed directly, we must leverage additional structure in order to estimate any causal effects. 

Our objective is to understand the model's structure and often apply it to tasks such as seeding. Thus, we rely on a correct model specification. The challenge with developing an IPW estimator arises when exposure is not observed. In such cases, it becomes impossible to determine which potential outcome was observed, violating the causal consistency assumption. Specifically, we don't know which potential outcome $Y_i$ represents (i.e., which exposure $v, Y_i = Y_i(v)$). 

\section{Additional Methodological Details} \label{sec:appendix: additional_methods}
In this section we discuss extensions to several aspects of the paper with respect to the paper. Before proceeding we also introduce the full statement of the generalized central limit theorem result which we use to derive our asymptotic results. 

\subsection{A central limit theorem for dependent data.} 
% In order to prove the asymptotics for a single network case we must utilize a central limit theorem (CLT) for dependent data; in particular, we leverage the version in  \citet{chandrasekhar2023general}. The conditions therein are implied by common assumptions of dependence such as $M$-dependence which includes  $\alpha$-,$\phi$- or $\rho$-mixing \citep{bradley2005basic}. We begin first with a definition of affinity sets which are used to motivate the theorem, then proceed with the CLT. 

Although network models do not neatly fit into conventional time series or spatial dependency categories, we provide a general framework by satisfying the necessary conditions through common dependence assumptions such as $M$-dependence. This includes scenarios characterized by $\alpha$-, $\phi$-, or $\rho$-mixing \citep{bradley2005basic}. Our approach begins by defining affinity sets, (sets for which there is high correlation with an outcome) that form the foundational framework for applying the CLT, setting the stage for demonstrating its relevance and utility in analyzing network data. 

\begin{definition}[Affinity sets]
    Denote a triangular array of mean $0$ random vectors $W_{1:n}^{(n)}$ with dimension $p$. Let $\mathcal{A}_{(i,d)}^{(n)}$ denote an affinity set which contains all of the variables in the triangular array which are highly correlated with $W^{(n)}_{i,d}$, the $d^{th}$ dimension of the $i^{th}$ random variable. 
\end{definition}
The affinity sets can be used to construct a matrix which contains the bulk of the covariance across observations and dimensions. The regularity conditions can be understood as control of the covariance within affinity sets \eqref{eq: within aff set covariance control}, control of the covariance across affinity sets \eqref{eq: cross aff set covariance control} and control of the covariance outside of the affinity sets \eqref{eq: outside aff set covariance control}. We collectively refer to these as the affinity set conditions. The affinity sets can be used to construct a covariance matrix $\Gamma_{n,dd'} = \sum_{i = 1}^{n}\sum_{(j,d') \in \mathcal{A}_{(i,d)}^{(n)}} \text{cov}(W^{(n)}_{i,d},W^{(n)}_{j,d'})$. 
\begin{align}
    \sum_{(i,d):(j,d'),(k,d'')}\E[W_{i,d}|W_{j,d'}W_{k,d''}] &= o(\norm{\Gamma_n}_{F}^{3/2}), \label{eq: within aff set covariance control} \\
    \sum_{(i,d),(j,d');(k,d''), (l,\hat d)}\text{cov}(W_{i,d}W_{k,d''},W_{j,d'}W_{l,\hat d} ) &= o(\norm{\Gamma_n}_{F}^{2}), \label{eq: cross aff set covariance control} \\
    \sum_{(i,d)}\E[|\vect{W}_{-i,d}\E[W_{i,d}\vect{W}_{-i,d}]|] &= o(\norm{\Gamma_n}_F). \label{eq: outside aff set covariance control} 
\end{align}

\begin{theorem}[Theorem 1 from \citet{chandrasekhar2023general}] \label{theorem: generalized CLT}
    Denote a mean $0$ triangular array of random vectors $W_{1:n}^{(n)}$. If a collection of affinity sets $\mathcal{A}_{(i,d)}^{(n)}$ satisfy the conditions of equations \eqref{eq: within aff set covariance control}, \eqref{eq: cross aff set covariance control} and \eqref{eq: outside aff set covariance control}. Then 
    $$ \Gamma_n^{-1/2} S_n \to_d N(0,I_p) $$
\end{theorem}
The authors illustrate several examples under which these conditions are sufficient for the this central limit theorem to hold. This theorem will be useful for proving our asymptotic results. 

\subsection{Inference for the OLS Estimator} \label{sec: OLS estimator inference}
Here we first give the full theorem and regularity conditions with respect to the  linear model. 
\begin{theorem} \label{theorem: ols asymptotics}
    Let $\tilde H_i(\theta) = \E[\tilde h(S_i(G), V_i(G))|\vect{a}, \vect{X}, G^*; \theta]$. The OLS estimator uses the model averaged coefficients $\tilde H_i(\theta)$ in place of the true unobserved coefficients $\tilde h_i$. Let $\mathsf{H}_n(\theta) = \frac{1}{n} \sum_{i = 1}^n \tilde H_i(\theta) \tilde H^T_i(\theta)$. Given an estimate of the model parameters $\hat \theta$,  we define the 
    \[
        \hat \beta_{lm} = \mathsf{H}^{-1}_n(\hat \theta) \frac{1}{n} \sum_{i = 1}^n \tilde H_i(\hat \theta) Y_i 
    \]
    Let $u_i = (\tilde h(S_i(G), V_i(G)) - \tilde H_i(\theta_0))\beta_0 + \epsilon_i.$  Suppose the following conditions hold for all $n$. \\
    \textbf{Model Regularity conditions}
    \begin{enumerate}[label=E\arabic*.,ref=E\arabic*]
    \item $\hat \theta$ is a $s(n)$-consistent estimate of the graph parameters $\norm{\hat \theta - \theta_0} = o_P(s(n))$ \label{assumption: ols Model Estimation Rate}
    \item $|\mathsf{H}_n(\theta) - \mathsf{H}_n(\theta')| \leq b_n(\vect{Z}) \norm{\theta - \theta'}$ where $b_n(\vect{Z}) = O_P(1)$ (that is, $b_n(\vect{Z})$ is stochastically bounded). \label{assumption: ols Hessian Smoothness}
    \item $\max_{i} \norm{\tilde H_i(\theta) - \tilde H_i(\theta')} \leq b_n(\vect{Z}) \norm{\theta - \theta'}$ \label{assumption: ols Feature Smoothness}
    \item $\norm{H_i(\theta)} \leq M < \infty$ \label{assumption: ols boundedness}
    \item $ \left| \frac{1}{n}\sum_{i = 1}^n |u_i| -  \frac{1}{n}\sum_{i = 1}^n \E[|u_i|] \right| = o_P(1)$ \label{assumption: ols bounded error}
    \end{enumerate}
    
    Lastly, let $\Gamma_n$ denote a matrix that satisfies the following central limit theorem for the estimating function \\
    \textbf{Central Limit Theorem}
    \begin{enumerate}[label=F\arabic*.,ref=F\arabic*]
    \item For the array of random variables $\mathcal{G}_i = \frac{1}{n}H_i(\theta_0) u_i$, there exists a set of affinity sets $\mathcal{A}_{(i,d)}^n$ such that \eqref{eq: within aff set covariance control}, \eqref{eq: cross aff set covariance control} and \eqref{eq: outside aff set covariance control} are satisfied with a corresponding matrix $\Gamma_n$, where $\sqrt{\lambda_{min}(\Gamma_n)} = r(n)$. \label{assumption: ols CLT of the estimating function}
    \end{enumerate}
    Then if $r(n) = o(s(n))$
    \[
        \Gamma_n^{-1/2} \mathsf{H}_n(\hat \theta) (\hat \beta_{lm} - \beta_{0}) \to_d N(0, I_p)
    \]
\end{theorem}

\subsection{Estimation of Network Models}
We next discuss the estimation of generative models of network formation using several datatypes. We summarize the information for using ARD in Table~\ref{tab: ARD rate of network model table} as discuss similar rates for other datatypes.

\begin{table}[htbp] \label{tab: ARD rate of network model table}
\footnotesize
\centering
\begin{tabular}{|l|c|c|}
\hline
Network Model &Norm   & ARD Rate                    \\ \hline
SBM           &$\sum_{k, k'}|{\hat P_{k k'} - P_{k k'}}|$  & $\tilde{O}_P(K/n)$              \\
Latent Space  &$\sup_{i \in \{1,2,\dots, n\}}|\hat \theta_i - \theta_i|$     & $O_P(\sqrt{log(n)/n})$ \\
Beta Model    &$\sup_{i \in \{1,2,\dots, n\}}|\hat \theta_i - \theta_i|$    & $O_P(\sqrt{log(n)/n})$ \\
Low-Rank Graphon &$\frac{1}{n^2}\norm{\hat \eta - \eta_0}_2$   & $\tilde{O}_P(1/T)$ \\
\hline
\end{tabular}
\caption{Summary of estimation rates with respect to model classes. The norms used for the latent space and beta models are with respect to their individual parameters $\theta_i$. We let $\eta_{0,ij} = P(G_{ij} = 1|\theta_0)$ denote the probability of two nodes connecting in the graphon model. Rates for the latent space and beta models are derived in \citet{breza2023consistently} and the low-rank graphon in \citet{alidaee2020recovering}. } 
\end{table}

\subsubsection{Estimation of the Stochastic Blockmodel Using Sampled Data} \label{sec: estimation of SBM}

We illustrate that it is possible to estimate the stochastic blockmodel using a diverse set of partial and sampled network data types. In each case, $\mathbf{P}_{kk'}$ refer to the cross-block probabilities, while $k_i \in \{1,2,\dots, K\}$ denote the node memberships. We consider \textit{partial network data} to be any subset of the network data which can be used to generate an estimate of the generative model $\widehat \theta$.

\begin{example}[Induced subgraph]
We sample $m \leq n$ of nodes in the graph randomly, with at least one node from each of the $K$ communities. Let $G'$ be the sub-graph induced by these $m$ nodes. Let $N'_k$ denote the set of sampled nodes in community $k$, assumed to be positive for each $k$. Let
\begin{equation*}
    \mathbf{\hat P}_{kk'} = \frac{1}{|N'_k| |N'_{k'}|} \sum_{i \in N'_k} \sum_{j \in G'_{k'}} G'_{ij} \;.
\end{equation*}
\end{example}

\begin{example}[Edges missing]
    Suppose that edges are missing according to some distribution. Let $G'$ be the observed graph, and suppose that $P(G'_{ij} = 1 | X_{ij} = x)$ is the probability of observing the edge $G'_{ij}$, given dyad-level covariates $X$ and the edge $G_{ij}$. Suppose that we have a consistent estimator of this conditional response. Then, 
    \begin{equation*}
        \mathbf{\hat P}_{kk'} = \frac{1}{|N'_k| |N'_{k'}|} \sum_{i \in N'_k} \sum_{j \in G'_{k'}} \frac{G'_{ij}}{\hat P(G'_{ij} = 1 | X_{ij})}\;.
    \end{equation*}  
\end{example}

\begin{lemma}[Rates for induced subgraph and Edges Missing] \label{lemma: Subgraph SBM rate}
    Consider an estimate for a stochastic blockmodel cross probabilities based on either the induced subgraph or the edges missing example of $m \leq n$. Let $m_k = |N_k| = \rho_k m$ for some $\rho_k \in (0,1)$. Then with probability at least $1 - \delta$ 
    \begin{equation} 
        |\mathbf{\hat P}_{kk'} - \mathbf{P}_{kk'} | \leq \frac{1}{\rho_{k}\rho_{k'} m}\sqrt{\frac{\log(2/\delta)}{2}} \label{eq: convergence rate}
    \end{equation}
    Further, suppose that $\sup_{x} |\hat P(G_{ij} = 1 | X_{ij} = x) - P(G_{ij} = 1 | X_{ij} = x)| = o_P(m^{-1})$ with $P(G_{ij} = 1 | X_{ij} = x) \geq \lambda > 0$. Then for large enough $m$, equation~\ref{eq: convergence rate} holds for the missing edges example as well. 
\end{lemma}

Lastly, we discuss respondent driven sampling. In this setting, community membership can be defined based on a partition of the covariates, thus allowing for an observable trait in the graph, a similar strategy is adopted by \citet{roch2018generalized}. 

\begin{example}[Respondent driven sampling]
    Let $i \in \{1,2,\dots, m\}$ denote the indices of a sample of individuals obtained through respondent driven sampling. An initial number of individuals are recruited as seeds, and subsequent individuals are recruited via referrals from the others in a population. ~\citet{tran2021estimation} develop a consistent estimator for the model parameters of the stochastic blockmodel. 

    Let $\tilde G_m$ be the subgraph of $G_n$ sampled from a set of nodes $\{1,2,\dots, m\}$. Let $M_{k}$ denote the number of individuals in the subsample of type $k$ and let $M^{\leftrightarrow}_{kk'}$ denote the number of connected individuals in the subgraph $\tilde G_m$. 

    The cross-type probabilities can be estimated as follows: 
    \begin{equation*}
        \vect{\hat P}_{kk'} = \begin{cases}
            \frac{M^{\leftrightarrow}_{kk'}}{M_{k}M_{k'}} \quad &\text{ When } k \not = k' \\
            \frac{M^{\leftrightarrow}_{kk}}{M_{k} (M_{k} - 1)} \quad &\text{ otherwise} 
        \end{cases}
    \end{equation*}
    ~\citet{tran2021estimation} illustrate the consistency of these parameters (Theorem 4.2 in their paper), in particular $|\vect{\hat P}_{kk'} - \vect{P}_{kk'}| = O_P(m^{-1})$ 
\end{example}

\subsubsection{Estimation of Other Network Models} \label{sec: estimation of Other Network Models}
Though we emphasise the estimation of the stochastic blockmodel, there are several other methods available for estimation of the network formation model. These include the beta model of \citet{chatterjeed2011}, in which the graph generation model consists of two model parameters $\nu_i, \nu_i$ possibly altered through some additional dyadic covariates $X^*_{ij}$

\begin{equation*}
    P(G_{ij} = 1|\theta_0) = \tilde f(\nu_i + \nu_j + \beta^T X^*_{ij})
\end{equation*}

where $\tilde f$ is a link function. Alternatively one can consider the latent space model of \citet{Hoff2002LatentAnalysis} which include latent positions on some unobserved manifold $\mathcal{M}^p$. 
\begin{equation*}
    P(G_{ij} = 1|\theta_0) = \tilde f(\nu_i + \nu_j + d_{\mathcal{M}^p}(Z_i, Z_j))
\end{equation*}
In each of these cases \citet{breza2023consistently} illustrate consistent estimation rates in the $\norm{\hat \theta - \theta_0}_{\infty} = \mathcal{O}_P\left(\sqrt{\frac{\log(n)}{n}}\right)$ with the use of aggregated relational data. Since this represents the coarsest datatype we expect similar rates to hold for subgraph sampling and respondent driven sampling. Though this rate is too slow for the to ignore the effect of the estimation of the graph model, in examples where one expect a high level of correlation among the outcomes it can be practical to use these methods.

\subsection{An EM algorithm for Logistic Regression} \label{sec: EM algorithm for Model Parameters}
Here we elaborate on the computation of a Z estimator. In general, an estimator may require specific implementation, we provide an illustrative example with logistic regression. Recall the characterization of the average estimating function $m_i(Y_i, \vect{a}, \vect{X}; \beta, \theta) = \E[\tilde m(Y_i, S_i(\vect{X}, G), V_i(\vect{a}, G); \beta)|\vect{Y}, \vect{a}, \vect{X}; \theta]$. Under this model, $P(Y_i = 1|S_i(\vect{X}, G), V_i(\vect{a}, G)) = \Lambda(\tilde h(S_i, V_i)^T \beta)$. 

In order to compute the new estimating function, we need to be able to consider the distribution of the graph, conditional on the observed outcome $Y_i$. Specifically. 
\al{
    P(G|Y_i, \vect{a}, \vect{X}, \beta, \theta) &= \frac{P(Y_i|G, \vect{a}, \vect{X}; \beta) P(G|\vect{a}, \vect{X}, \theta)}{P(Y_i|\vect{a}, \vect{X}, \beta, \theta)} \\
    &= \frac{P(Y_i|S_i(\vect{X}, G), V_i(\vect{a}, G); \beta) P(G|\theta)}{P(Y_i|\vect{a}, \vect{X}, \beta, \theta)}
}

In a standard missing data problem, one would impute the missing covariates directly, however, due to the dependence through the graph, this can be very difficult to achieve in practice. However, it will be straightforward to sample from the graph model $P(G|\theta)$. Using a simple approach, we can compute the maximizer exploiting standard software methods using an EM algorithm \citep{dempster1977maximum, wu1983convergence}. Suppose that we draw a sample of graphs from the generative model $\{G^{(l)}\}_{l = 1}^L \sim_{iid} P(G|\theta)$. 

Let $w_i(Y_i, G; \beta)$ define the weight of an observation. 
\al{
    w(Y_i, G; \beta) &= \frac{P(Y_i|S_i(\vect{X}, G), V_i(\vect{a}, G); \beta)}{P(Y_i|\vect{a}, \vect{X}, \beta, \theta)} \\
    &\approx \frac{P(Y_i|S_i(\vect{X}, G), V_i(\vect{a}, G); \beta)}{\frac{1}{L}\sum_{l = 1}^L  P(Y_i|S_i(\vect{X}, G^{(l)}), V_i(\vect{a}, G^{(l)}); \beta)}
}
We next construct the EM algorithm as follows. 

\begin{algorithm}
\begin{algorithmic}[1]
    \State Sample $\{G^{(l)}\}_{l = 1}^L \sim_{\text{iid}} P(G|\hat{\theta})$ denote a sample from the graph model and initialize parameters $\hat{\beta}^{(0)}$
    \For{$t \in \{1, 2, \dots, T\}$}
        \State \textbf{(E-step)} Compute the weighted empirical estimating function 
        $$m^{(t)}_n(\vect{Y}|\vect{a}, \vect{X}, \beta, \hat{\theta}) = \frac{1}{L}\frac{1}{n} \sum_{l = 1}^L \sum_{i = 1}^n\tilde{m}(Y_i, S_i(\vect{X}, G^{(l)}), V_i(\vect{a}, G^{(l)}); \beta)w(Y_i, G^{(l)}; \hat{\beta}^{(t - 1)})$$
        \State \textbf{(M-step)} Solve the new estimating function by solving: 
        $$m^{(t)}_n(\vect{Y}|\vect{a}, \vect{X}, \hat{\beta}^{(t)}, \hat{\theta}) = 0$$
    \EndFor
\end{algorithmic}
\end{algorithm}

In practice, this allows for one to use standard solvers for the (M-step), after sampling a single time with the (E-step). 

Additionally, one can include correlations across the observations $Y_i$ through the use of a generalized estimating equation approach. In other generalized linear models, additional assumptions may be required in order to model the full conditional distribution $P(Y_i|S_i(\vect{X}, G), V_i(\vect{a}, G); \beta)$ such as a dispersion component.

\subsection{Plug-in estimates of the Causal parameter} \label{sec: plug-in estimates} 

For many problems, the parameter of interest is a causal query conditional on the complete graph $G$ as described in section~\ref{sec: nonparamtericID}. For example, one may care about the expected number of adoptions after seeding an individual in block $k$ v.s. block $k'$. In this section, we illustrate how to construct an estimate of the causal parameter $\Psi(\vect{a}| G)$ using our conditional model estimation procedure. 

Let $\Psi(\vect{a}|\theta_0) = \E[ \Psi(\vect{a}|G)| \vect{a}, \vect{X}, \theta_0]$ be the average causal effect of policy $\vect{a}$ over all draws of the graph model $\theta_0$. We will establish conditions under which these two quantities are close to one another. 

Recall the true conditional mean function $\E[Y|S_i = s, V_i = v] = h_0(s,v)$. Under a correctly specified conditional model, $h_0(s,v) = h(s,v; \beta_0)$, and $\Psi(\vect{a}|\theta_0) = \Psi(\vect{a}|\beta_0, \theta_0)$ where 
\begin{equation}
    \Psi(\vect{a}|\beta, \theta) = \frac{1}{n} \sum_{i = 1}^n \E[f(V_i, S_i; \beta)| \vect{a}, \vect{X}, \theta]. 
\end{equation}
In order to estimate $\Psi(\vect{a}| G)$ we plug-in the estimates for the mean model and network model $\Psi(\vect{a}|\hat \beta, \hat \theta)$. We next discuss the asymptotics of the plug-in estimate. 
\begin{lemma}[Inference for a plug-in causal parameter] \label{lemma: plug in inference}
Assume the conditions of \ref{assumption: Z estimator regularity conditions}. 
Further, assume: 
\begin{equation}
    \sup_{\beta}\left|\E[h(S_i(\vect{X}; G), V_i(\vect{a}| G); \beta)|\vect{a}, \vect{X}, \theta] - \E[h(S_i(\vect{X}; G), V_i(\vect{a}| G); \beta)|\vect{a}, \vect{X}, \theta']\right| \leq b_i\norm{\theta - \theta'}  \label{assumption: smooth in parameter} 
\end{equation}
where $b_i \leq M < \infty$. Denote $$Q_n(\beta) := \frac{1}{n}\sum_{i = 1}^n \frac{\partial}{\partial \beta'} \E[h(S_i(\vect{X}; G), V_i(\vect{a}| G); \beta')|\vect{a}, \vect{X}, \theta_0]\bigg|_{\beta' = \beta} \in \mathbb{R}^{1 \times p}$$ and 
$$\tilde \omega_n := Q_n(\beta_0)D_n(\beta_0)\Gamma_n D_n(\beta_0)^{T}Q_n(\beta_0)^{T}.$$
If $s(n) = o(\sqrt{\tilde \omega_n})$. Then 
\begin{equation}
    \tilde \omega_n^{-1/2}(\Psi(\hat \beta, \hat \theta) - \Psi( \beta_0, \theta)) \to_d N(0,1)
\end{equation}
\end{lemma}

This lemma is essentially an application of the delta method, with the additional caveat that we estimate $\theta$ before the plug-in estimate. As before, this requires a fast estimate of the graph generative model parameter, but we add the slightly different assumption (\cref{assumption: smooth in parameter}) that the smoothness in the model class is over the conditional response models $\E[h(S_i, V_i; \beta) | \theta]$, rather than the estimating function $\tilde m(Y,S,V|\beta,\theta)$. 

\textbf{Convergence of the causal parameter to the average over graphs} \label{sec: parameter convergence}

As we have previously discussed, we can only hope to estimate $\Psi(\vect{a}|\theta_0)$ as we do not have access to the full graph $G$. We next introduce a simple conditions under which the parameter $\Psi(\vect{a}| G)$ is close to its average over draws of the graph $G \sim \theta_0$, $\Psi(\vect{a}| \theta_0)$. 

\begin{assumption}[$v_n$-response dependence] \label{assumption: average parameter graph dependence}
    For any graph draw $G$ let $G^{'(ij)}$ denote the graph $G$ with the ${ij}$ entry swapped from $0$ to $1$ or vice versa. Let $c_{ij, n}$ denote the bounds of the differences such that 
    \begin{equation}
	   \left| \frac{1}{n} \sum_{i = 1}^n h_0(S_i(\vect{X}, G), V_i(\vect{a}, G)) - h_0(S_i(\vect{X}, G^{'(ij)}), V_i(\vect{a}, G^{'(ij)}))  \right| \leq c_{ij, n}
    \end{equation}
    And let $v_n^2 = \sum_{ij: i \not = j} c^2_{ij, n}$
\end{assumption}

\begin{lemma} \label{lemma: average parameter graph sample}
    Under Assumption \ref{assumption: average parameter graph dependence} 
    \begin{equation*}
        \Psi(\vect{a}|G) - \Psi(\vect{a}|\theta_0) = O_P(v_n)
    \end{equation*}
\end{lemma}
The proof is a one-line application of McDiarmid's inequality. Previous related work such as \citet{breza2023consistently} typically assume that such a quantity is consistent, however here we quantify the rate here. We next highlight an example; 

\begin{example}[Conditional Mean Function Example]
    We abbreviate $G = G$ and $G' = G^{'(kl)}$. Let $h_0(S_i(\vect{X}, G), V_i(\vect{a}, G)) = \beta_0 + \beta_1 a_i + \beta_2 X_i + \beta_3 \sum_{l \not = i} \frac{X_lG_{kl}}{n} + \beta_4 \sum_{l \not = k} \frac{a_lG_{il}}{n}$ denote a linear response function dependent on the density of connected neighbors. Suppose that the covariate values are bounded $|X_i| \leq M < \infty$. Then: 
    \al{
        &\left|\frac{1}{n}\sum_{k = 1}^n h_0(S_i(\vect{X}, G), V_i(\vect{a}, G)) - h_0(S_i(\vect{X}, G'), V_i(\vect{a}, G'))\right| \\
        &= \left|\frac{1}{n}\sum_{k = 1}^n \sum_{l \not = k} \beta_3\frac{ X_l (G_{kl} - G'_{kl})}{n} +  \beta_4\frac{a_l (G_{kl} - G'_{kl})}{n}\right| \\
        &\leq \left|\frac{1}{n}\frac{X_j + X_i}{n} +  \frac{|a_j| + |a_i| }{n}\right| \\
        &\leq (2M + 2)/n^2
    }
    Applying \ref{lemma: average parameter graph sample} illustrates that: $\Psi(\vect{a}|G) - \Psi(\vect{a}|\theta_0) = O_p(n^{-1})$. Hence in order to estimate the expected average outcome, all we need is a consistent estimate of the model parameters $\beta_0$. 
\end{example}

\subsection{Optimal design for a $Z$-estimator}
Here we illustrate the optimal design approach for Z-estimators. In this example, the variance itself may depend on the a parameter $\beta$, and thus one can include a working candidate for the parameter $\beta'$.  In general, one could also propose a feasible range of the working parameters $\beta'$ and consider the worst case variance in that range. 

\begin{algorithm}[H]
\caption{Saturation Randomized Design Variance.}
\begin{algorithmic}[1]
    \State \textbf{Inputs: } Working covariance $\Gamma_n$, model estimate $\hat \theta$, working parameter $\beta'$ 
    \State Sample $L$ draws from the graph model $\{\hat{G}^{(l)}\}_{l = 1}^L \sim \hat \theta$ 
    \State Sample $R$ treatments $\{\vect{a}_{r}\}_{r = 1}^R$ according to the block-saturation levels $\vect{\tau}$. 
    \For{$r \gets 1$ \textbf{to} $R$} 
    \State Compute $ \hat D_{r}(\vect{a}) = \frac{1}{nL}\sum_{l = 1}^L \sum_{i = 1}^n \nabla_{\beta} m_i(Y_i, S_i V_i ; \hat G^{(l)}, \beta') $ 
    \State Compute the variance for a single draw $\vect{a}_r$: 
    $$\upsilon^{\phi}(\vect{a}_r; \hat \theta) = \phi^T\hat D_{r}(\vect{a})^{-1} \Gamma_n \hat D_{r}(\vect{a})^{-1 T} \phi $$
    \EndFor
    \State Average over each of the draws $\bar{\upsilon}(\vect{\tau}; \hat \theta) = \sum_{r = 1}^R \upsilon^{\phi}(\vect{a}_r; \hat \theta)$
\end{algorithmic} 
\label{alg: Saturated Random Design Variance Z Estimator}
\end{algorithm}

\subsection{Experimental Design Variance Minimization With Model Uncertainty} \label{sec: Variance Minimization with Model Uncertainty}
As an extension of our variance minimizing procedure, we can incorporate the uncertainty in our estimates of the model parameters. For instance, consider the following parametric bootstrap approach for estimating the model parameters of the stochastic blockmodel when using ARD. 

For example, consider a scenario where we utilize the stochastic blockmodel and we collect ARD. Denote $\hat \theta = (\{\hat Z_i\}_{i = 1}^n, \vect{\hat P})$ the initial estimate of the model as computed from Lemma~\ref{lemma: ARD_clustering}. We can construct a sampling distribution of $\hat \theta^{(b)}$ using the following procedure. Let $X^*_{it}$ denote the ARD responses of the number of connections individual $i$ has to someone of trait $t$ and let $T_i \in \{0,1\}^T$ denote the trait memberships of the corresponding individuals. 
\begin{algorithm}[H]
\caption{Bootstrap ARD algorithm.}
\begin{algorithmic}[1]
    \State Estimate $\hat{\theta}$ from $\vect{X}^*$
    \For{$b \in \{1, 2, \dots, B\}$}
        \State Sample $G^{(b)} \sim \hat{\theta}$
        \State Construct the ARD vector based on the resampled responses $X^{*(b)}_{it}$ using counts according to connections of $G^{(b)}$ to the nodes with corresponding traits $\{T_i\}_{i=1}^n$
        \State Estimate $\hat{\theta}$ from $\vect{X}^{*(b)}$
    \EndFor
\end{algorithmic}
\label{alg: bootstrap ard}
\end{algorithm}

% \begin{enumerate}
%     \item Estimate $\hat \theta$ from $\vect{X}^*$ 
%     \item For $b \in \{1,2,\dots, B\}$
%     \begin{enumerate}
%         \item Sample $G^{(b)} \sim \hat \theta$
%         \item Construct the ARD vector based on the resampled responses $X^{*(b)}_{it}$ using counts according to connections of $G^{(b)}$ to the nodes with corresponding traits $\{T_i\}_{i= 1}^n$ \label{step: }
%         \item Estimate $\hat \theta$ from $\vect{ X}^{*(b)}$ 
%     \end{enumerate}
% \end{enumerate}
% Double check this environment

% Additionally one can scale $X^{*(b)}_{it}$ by $\frac{n}{m}$ if one only uses a sub-sample of the whole graph using ARD. 

This approach can work for any procedures which can allow for a sampling distribution of the model parameters $\{\hat \theta^{(b)}\}_{b = 1}^B$. For example \citet{baraff2016estimating} considers a nonparametric bootstrap for respondent driven sampling. 

In all such cases we would like to include thee uncertainty in $\hat \theta$ to the saturation assignment, we apply Algorithm~\ref{alg: Saturated Random Design Variance OLS} (or Algorithm~\ref{alg: Saturated Random Design Variance Z Estimator}) to each of the $b$ draws. Using the distribution of variances obtained over the $b$ draws, one can compute average or upper confidence bounds on the variance.  For example in the simulation illustrated in section~\ref{sec: example Local Diffusion} we select saturations based on the 2-standard deviation upper confidence bound of the average variance across $b \in \{1,2,\dots, B\}$. 

\section{Proofs of Theorems} \label{sec:appendix: theorem_proofs}

In this section, we introduce the proofs for the results in the main paper as well as the additional theoretical results presented in section~\ref{sec:appendix: additional_methods}.

\subsection{Proof of Lemma~\ref{lemma: ARD_clustering}}

\begin{proof}
    Under the stochastic blockmodel assumption, the true latent traits are some discrete type $k_i \in \{1,2,\dots, K\}$. Then the mean connection probability $Z_{ck}$ is simply a mixture over the connection probabilities, weighted by $P(k_j = k'|t_j = t)$. Let $N_k$ denote the set of individuals with group $k$ membership. Furthermore, let $n = |N_k|$. Denote analogous quantities for the trait memberships $N_t$ as well as the intersection of $k$ and $t$ by $N_{kt}$. When we have a correct clustering. Denote $\hat P_{k t} = \frac{1}{n_k} \sum_{i \in N_k} \frac{1}{n}\tilde Y_{i t}$. Assuming independent samples conditional on the graph clusters, let $P_{k t} = \frac{1}{n_{t}} \sum_{k' \in [K]} \sum_{i \in N_{tk'}} P_{k k'}$ denote the mean probability of connection averaged over the clusters conditional on their latent traits. Let $\omega_{kt} = \frac{n_{kt}}{n_t}$. 

    We can express $\tilde P_{k t} = P(G_{ij} = 1 | k_i = k, t_j = t)$ as a weighted sum of the connection probabilities from the constituent distributions. If the true clusters are known, then these proportions $\omega_{kt}$ are known exactly from the data. 
    Then 
    \al{
         \tilde P_{k t} &= P(G_{ij} = 1 | k_i = k, t_j = t) \\
         &= \sum_{k' = 1}^KP(G_{ij} = 1 | k_i = k, k_j = k', t_j = t) P(k_j = k'|t_j = t) \\
         &= \sum_{k' = 1}^KP(G_{ij} = 1 | k_i = k, k_j = k', t_j = t) \omega_{k't} \\
         &= \sum_{k' = 1}^KP(G_{ij} = 1 | k_i = k, k_j = k') \omega_{k't} \\
         &= \sum_{k' = 1}^KP_{kk'} \omega_{k't} 
    }
    Expressing this relationship over the whole set of matrices, we have: 
    \al{
        \tilde P = \Omega  P
    }
    Where $\Omega_{tk} = \frac{n_{tk}}{n_k}$.  We can solve this system as long as the columns of $\Omega$ are linearly independent. Therefore: 
    \al{
        P = (\Omega^T \Omega)^{-1} \Omega^T \tilde P
    }

    We next bound the estimation error in Frobenius norm of the true cross-cluster probabilities
    \al{
        \norm{\hat P - P}_F &= \norm{(\Omega^T \Omega)^{-1} \Omega^T(\hat{\tilde{P}} - \tilde{P})}_F \\
        &\leq \norm{(\Omega^T \Omega)^{-1} \Omega^T}_F \norm{(\hat{\tilde{P}} - \tilde{P})}_F \\
        &\leq \sqrt{\norm{(\Omega^T \Omega)^{-1} \Omega^T}^2_F } \norm{(\hat{\tilde{P}} - \tilde{P})}_F \\
        &\leq \sqrt{\text{Tr}\left((\Omega^T \Omega)^{-1} \Omega^T \Omega (\Omega^T \Omega)^{-1} \right) } \norm{(\hat{\tilde{P}} - \tilde{P})}_F \\
        &= \sqrt{\text{Tr}\left((\Omega^T \Omega)^{-1}\right) } \norm{(\hat{\tilde{P}} - \tilde{P})}_F \\
    }
    Since we assume that $\Omega$'s column's are linearly independent, then $\Omega^T\Omega$ is invertible. Therefore, what remains is bounding the Frobenius norm of $\norm{(\hat{\tilde{P}} - \tilde{P})}_F$. 

    For each element, let 
    \al{
        \hat{\tilde{P}}_{tk} &= \frac{1}{n_kn_t} \sum_{i \in N_k} \tilde Y_{ik} \\
        &= \frac{1}{n_kn_t} \sum_{i \in N_k} \sum_{j \in N_t} G_{ij} 
    }

    Therefore, applying Hoeffding's inequality
    \al{
        P( |\hat{\tilde{P}}_{tk} - \tilde{P}_{tk}| \geq \epsilon ) &\leq 2 \exp \left( -2 \epsilon^2 n_k n_t\right)
    }
    Letting $\rho_k = \frac{n_k}{n}, \rho_t = \frac{n_t}{n}$, then 
    \al{
        P( |\hat{\tilde{P}}_{tk} - \tilde{P}_{tk}| \geq \epsilon ) &\leq 2 \exp \left( -2 \epsilon^2 \rho_k \rho_t n^2\right)
    }
    Therefore, by a union bound, 
    \al{
        P( \max_{k,t}|\hat{\tilde{P}}_{tk} - \tilde{P}_{tk}| \geq \epsilon ) &\leq 2KT \exp \left( -2 \epsilon^2 \rho_k \rho_t n^2\right) \\
        \implies P( \sum_{k,t}|\hat{\tilde{P}}_{tk} - \tilde{P}_{tk}| \geq KT\epsilon ) &\leq 2KT \exp \left( -2 (KT)^2\epsilon^2 \rho_k \rho_t n^2\right)
    }
    Therefore, 
    \al{
        \norm{\hat{\tilde{P}}_{tk} - \tilde{P}_{tk}}_1 &= \OO_P(\frac{KT\sqrt{\log(KT)}}{n })
    }

    Hence 
    \al{
        \norm{\hat P - P}_2 &= \OO_P(\frac{KT\sqrt{\log(KT)}}{n})
    }

    Lastly, we show that as $n$ grows, the probability of achieving a correct clustering of the true block memberships approaches $1$. Recall that $n_t = \rho_t n$, and let $\underline{\rho_T} = \min_{t} \rho_t$. By Hoeffding's inequality: $P(\norm{X^\dagger_i - Z_{k_i}} > \epsilon_n) \leq 2\exp(-2 \epsilon^2_n n/\underline{\rho_T})$. Taking a union bound over all response vectors, $P(\max_{i}\norm{X^\dagger_i - Z_{k_i}} > \epsilon_n) \leq 2n 2\exp(-2 \epsilon^2_n n/) \to 0$ for all $\epsilon_n = o(\sqrt{\log(n)/n})$. 
    
    Therefore, as $n$ grows, the normalized response vectors in each cluster become well separated, and once $\epsilon_n < \min \norm {Z_{k} - Z_{k'}}/2$, then all clusters will be well separated and naively hierarchical agglomerative clustering will consistently group the blocks together for $K$ clusters. Therefore for example, if we let $\epsilon_n = \log(n) n^{-1/2}$,then $P( \text{max}_{i} \{\hat k \not =  k \} = O(\frac{1}{n}) )$. Of course the labels learned are only consistent up to permutation. We exploit the fact that as referred to in \citet{breza2023consistently}, the clustering problem gets easier as the sample size grows. Let $\mathcal{E}$ be the event that $\hat k_i = k_i$ up to permutation for all $i \in \{1,2,\dots, n\}$, i.e. $P(\max_{i} |\hat k_i = k_i|  > 0 ) = 1 - P(\mathcal{E}) \leq \frac{1}{n}$. Since the estimators are not necessarily independent of the event of perfect classification. 

    \al{
        P(\norm{\hat P_{\vect{\hat k}} - P_{\vect{k}}} > \epsilon) 
        &= P(\norm{\hat P_{\vect{\hat k}} - P_{\vect{k}}} > \epsilon | \mathcal{E})P( \mathcal{E}) +  P(\norm{\hat P_{\vect{\hat k}} - P_{\vect{k}}} > \epsilon | \mathcal{E}^c)P( \mathcal{E}^c) \\
        &\leq P(\norm{\hat P_{\vect{\hat k}} - P_{\vect{k}}} > \epsilon , \mathcal{E}) + P( \mathcal{E}^c) \\
        &\leq P(\norm{\hat P_{\vect{\hat k}} - P_{\vect{k}}} > \epsilon , \mathcal{E}) + \frac{1}{n} \\
        &= P(\norm{\hat P_{\vect{k}} - P_{\vect{k}}} > \epsilon , \mathcal{E}) + \frac{1}{n} \text{ Since } \mathcal{E} \text{ indicates the correct classification}\\
        &\leq P(\norm{\hat P_{\vect{k}} - P_{\vect{k}}} > \epsilon) + \frac{1}{n} \\
        &\leq \sqrt{\text{Tr}\left((\Omega^T \Omega)^{-1}\right) } 2KT \exp \left( -2 (KT)^2\epsilon^2 \rho_k \rho_t n^2\right) + \frac{1}{n} 
    }
    Therefore 
    $$ \norm{\hat P_{\vect{\hat k}} - P_{\vect{k}}} = \OO_P(\frac{KT\sqrt{\log(KT)}}{n })$$
\end{proof}

\subsection{Proof of \cref{theorem: Z estimator asymptotics}}
\begin{proof}
    We emphasise that in general, the outcomes $\vect{Y}$ may be dependent, and this is reflected through correlation in the estimating functions (or the residuals in the case of OLS).
    We will partition the proof into two sections. First, we will prove the consistency of the estimator $\hat \beta $ and secondly, we will prove the central limit theorem. \\
    \textbf{Consistency: } 
    The following result hinges on a typical consistency proof for the M or Z estimators using a structure similar to those found in Chapter 5 of \citet{Vaart1998AsymptoticStatistics}. 
    First, we denote that: \al{
        m_n(\vect{Z}; \hat \beta, \hat \theta) - g_n(\vect{Z}; \hat \beta) &= m_n(\vect{Z}; \hat \beta, \hat \theta) - m_n(\vect{Z}; \hat \beta, \theta_0) \\
        &\leq b_n(\vect{Z})\norm{\hat \theta - \theta_0} \\
        &= O_P(1)o_P(s(n)) \\
        &= o_P(s(n))
    }
    Next, we can see that, based on this expansion,  
    \al{
        m_n(\vect{Z}; \hat \beta, \hat \theta) &= 0 \\
        \implies 0 &= (m_n(\vect{Z}, \hat \beta, \hat \theta) - g_n(\vect{Z}, \hat \beta)) + g_n(\vect{Z}; \hat \beta)  \\
        &= o_P(s(n)) + g_n(\vect{Z}; \hat \beta)  \text{ By  \ref{assumption: Uniform LLN}}
    }
    At this point, we can now treat this as a standard Z-estimation problem. Therefore, by \ref{assumption: Uniform LLN} and \ref{assumption: separatedness of solution}, then $\hat \beta$ is a solution to the estimating function $g$ and is therefore consistent by an application of Theorem 5.9 of \citet{Vaart1998AsymptoticStatistics}. \\
    
    \textbf{Asymptotic Normality: } We illustrate asymptotic normality through a Taylor series expansion argument. As we saw in the consistency part of the proof
    \al{
         g_n(\vect{Z}; \hat \beta) = o_P(s(n)) \\
    }
    For brevity in notation, we suppress the dependence on $\vect{Z}$, which is implicit for functions, with the subscript $n$. Using a Taylor expansion around $\beta_0$, and let $\tilde \beta_j \in [\beta_{0,j}, \hat \beta_{j}]$ for $ \beta_{0,j}\leq  \hat \beta_{j}$ and $\tilde \beta_j \in [\hat \beta_{j}, \beta_{0,j}]$ otherwise. 
    \al{
        g_n(\hat \beta) &= g_n(\beta_0) +  D_n(\vect{Z};\beta_0) (\hat \beta - \beta_0) +  \sum_{jk} \frac{\partial^2}{\partial \beta_j \beta_k} g_n(\vect{Z}; \tilde \beta)(\hat \beta_j - \beta_{0,j})(\hat \beta_k - \beta_{0,k}) \\
        &=  g_n(\beta_0) + D_n(\vect{Z};\beta_0) (\hat \beta - \beta_0) + o_P(s(n) + \norm{\hat \beta -\beta_0})
    }
    by the application of the consistency and \ref{assumption: Uniform LLN}. Therefore, we focus on main terms. By Assumption \ref{assumption: CLT of the estimating function}. 

    Therefore: 
    \al{
        \Gamma_n^{-1/2}D_n(\vect{Z};\beta_0)(\hat \beta - \beta_0) = \Gamma_n^{-1/2} g_n(\beta_0) + o_p(\frac{s(n)}{r(n)})
    }
    
    Noting that $D_n(\beta_0) - D(\beta_0)= o_P(1)$, by an application of Slutsky's lemma: 
    \al{
         \Gamma_n^{-1/2} D(\beta_0)(\hat \beta - \beta_0) &\to_d N(0, I_p)
    }
    and therefore, the proof is complete.
\end{proof}

\subsection{Proof of Lemma~\ref{lemma: graph misspecification}}
We first include a useful lemma for bounding the approximation of the error of the graphon model. 
\begin{lemma}[Lemma 2.1 of \citet{gao2015rate}]\label{lem:graphon_rate}
    Denote $k_i \in \{1,2,\dots, K\}$ are the block memberships of a stochastic-blockmodel with average connection probabilities across blocks $\bar{\eta}_{ij} = P_{k,k'} = \frac{1}{n_k n_{k'}} \sum_{i, j : k_i = k, k_j =  Z_i}\sum_{l : Z_l = Z_j} \eta_{kl}$. If the true graphon $g \in \mathcal{H}_\alpha(M)$, then, there exists some membership vector $\vect{k}$ and corresponding average across block probabilities $P_0$ such that:
    \al{
        \frac{1}{n^2} \sum_{ij} (\eta_{ij} - \bar{\eta}_{ij})^2 &\leq M^2 \left( \frac{1}{K^2}\right)^{\alpha \wedge 1 }
    }
\end{lemma}

We now proceed with a the proof of the lemma. 
\begin{proof}
    We firstly use a Taylor expansion of $L_n(\beta_0,\eta_*)$ where $\tilde \beta$ is an element-wise intermediate value of $\beta$ and $\tilde \beta$
    \al{
        L_n(\beta_0,\eta_*) &= L_n(\beta_*,\eta_*) + \frac{\partial }{\partial \beta} L_n(\beta,\eta_*) \bigg |_{\beta = \beta_*}(\beta_0 - \beta_*) \\
        &+ \sum_{jk} \frac{\partial^2 }{\partial \beta_j \partial \beta_k} L_n(\tilde \beta,\eta_*) (\beta_{0j} - \beta_{*j})(\beta_{0k} - \beta_{*k})\\
        \frac{\partial }{\partial \beta} L_n(\beta,\eta_*) \bigg |_{\beta = \beta_*}&(\beta_0 - \beta_*) = - L_n(\beta_0,\eta_*) + 
        \sum_{jk} \frac{\partial^2 }{\partial \beta_j \partial \beta_k} L_n(\tilde \beta,\eta_*)(\beta_{0j} - \beta_{*j})(\beta_{0k} - \beta_{*k}) 
    }
    Since we assume $L_n(\beta, \eta_*)$ is twice continuously differentiable in $\beta$, and $\mathcal{B}$ is compact, then $\frac{\partial^2 }{\partial \beta_j \partial \beta_k} L_n(\tilde \beta,\eta_*)$ is bounded. Therefore, 
    \al{
        \norm{\beta_0 - \beta_*}_2 &\leq \frac{|L_n(\beta_0,\eta_*)|}{\lambda\sqrt{p}} + O(\norm{\beta_0 - \beta_*}_2^2)
    }
    Lastly, by our continuity assumptions, $|L_n(\beta_0,\eta_*)| \leq L \norm{\eta_0 - \eta_*}_2/n \leq LM K^{- (\alpha \wedge 1)}$. After applying this, our proof is complete. 
    
\end{proof}

%%%% Begin: TODO: Shuffle reordering afterwards 

\subsection{Proof of Lemma~\ref{lemma: Subgraph SBM rate}}
\begin{proof}
    The proof is straightforward application of Hoeffding's inequality. Given an $m$ node subsample of the full graph, and given their known types. Since $ \mathbf{\hat P}_{kk'} = \frac{1}{\rho_{kk'}m}\sum_{i,j}G_{ij}I(k_i = k, k_j = k')$, then the final result is a direct application of Hoeffding's inequality. 

    For the missing data case, we can plug-in the estimate of the edge sampling $P(G_{ij} = 1 | X_{ij} = x)$ in order to correct for the missingness of the edges. If $\sup_{x} |\hat P(G_{ij} = 1 | X_{ij} = x) - P(G_{ij} = 1 | X_{ij} = x)| = o_P(m^{-1})$ then the estimation of the propensity is negligible and we can correct for the missingness of edges. 
\end{proof}

\subsection{Proof of \cref{theorem: ols asymptotics}}
\begin{proof}
    We first we expand the form of the OLS estimator. 
    \begin{align*}
        \hat \beta_{lm} &= \mathsf{H}^{-1}_n(\hat \theta) \frac{1}{n} \sum_{i = 1}^n \tilde H_i(\hat \theta) Y_i \\
        &= \mathsf{H}^{-1}_n(\hat \theta) \frac{1}{n} \sum_{i = 1}^n \tilde H_i(\hat \theta)  (\tilde H^T_i(\theta_0) \beta_0 + u_i) \\
        &= \mathsf{H}^{-1}_n(\hat \theta) \frac{1}{n} \sum_{i = 1}^n \tilde H_i(\hat \theta)  (\tilde H^T_i(\hat \theta) \beta_0 +  (\tilde H^T_i(\theta_0) - \tilde H^T_i(\hat \theta))\beta_0 + u_i) \\
        &= \beta_0 + \underbrace{\mathsf{H}^{-1}_n(\hat \theta) \frac{1}{n} \sum_{i = 1}^n \tilde H_i(\hat \theta) (\tilde H^T_i(\theta_0) - \tilde H^T_i(\hat \theta))\beta_0}_{(A)}\\
        &+ \underbrace{\mathsf{H}^{-1}_n(\hat \theta) \frac{1}{n}\sum_{i = 1}^n  (\tilde H_i(\hat \theta) - \tilde H_i(\theta_0)) u_i}_{(B)} + \underbrace{\mathsf{H}^{-1}_n(\hat \theta) \frac{1}{n}\sum_{i = 1}^n \tilde H_i( \theta_0) u_i}_{(C)}
    \end{align*}

    We next bound terms $(A)$ and $(B)$ after which, the asymptotic distribution of $(C)$ will be apparent. 
    
    We note that the Hessian evaluated at the true model parameters can be evaluated $\mathsf{H}_n(\hat \theta) = \mathsf{H}_n(\theta_0) + o_P(s(n))$ by assumptions \ref{assumption: ols Hessian Smoothness} and \ref{assumption: ols Model Estimation Rate}. By the continuous mapping theorem $\mathsf{H}_n(\hat \theta) = \mathsf{H}_n(\theta_0) + o_P(s(n))$. We see that $(A) = o_P(s(n))$ by assumptions~\ref{assumption: ols boundedness}, \ref{assumption: ols Hessian Smoothness} and \ref{assumption: ols Model Estimation Rate}. 
    Next, by the stochastic boundedness of the error \ref{assumption: ols bounded error} and applying H\"{o}lder's inequality. 
    \al{
        \frac{1}{n}\sum_{i = 1}^n  (\tilde H_i(\hat \theta) - \tilde H_i(\theta_0)) &\leq (\frac{1}{n}\sum_{i = 1}^n |u_i|) \max_{i} \norm{\tilde H_i(\hat \theta) - \tilde H_i(\theta_0)} \\
        &= o_P(s(n))
    }
    Therefore: 
    \al{
        \Gamma_n^{-1/2}  \mathsf{H}_n(\hat \theta) (\hat \beta_{lm} - \beta_0) = \Gamma_n^{-1/2} \sum_{i = 1}^n \tilde H_i(\theta_0) u_i + o_P\left(\frac{s(n)}{r(n)}\right) \to_d N(0, I_p)
    }
    by \ref{assumption: ols CLT of the estimating function} and Slutsky's Lemma, completing the proof. 
\end{proof}

\subsection{Proof of Lemma~\ref{lemma: plug in inference}}
\begin{proof}
    The proof follows from an application of the delta method, with the additional caveat that we must account for the estimation of the model parameters $\theta_0$
    In this case: 
    \al{
        |\Psi(\hat \beta, \hat \theta) - \Psi(\hat \beta, \theta_0) | &\leq \frac{1}{n}\sum_{i = 1}^n b_i \norm{\hat \theta - \theta_0} \\
        &= o_P(s(n))
    }
    The remainder of the proof follows from a simple application of the delta method using the plug-in estimator $\Psi(\hat \beta, \theta_0)$. See Theorem 3.1 of \citet{Vaart1998AsymptoticStatistics}. 
\end{proof}

%%%% END: Shuffle reordering 

\section{Additional Simulations}

Here we provide additional details with respect to several aspects of our methodology.  We also include further details on several of the details for the implementation of competing methods in section \ref{sec: Experimental Details}. 

\subsection{Coverage of the GATE}
\begin{figure}[htb!]
    \centering
    \includegraphics[height = 0.4\textheight]{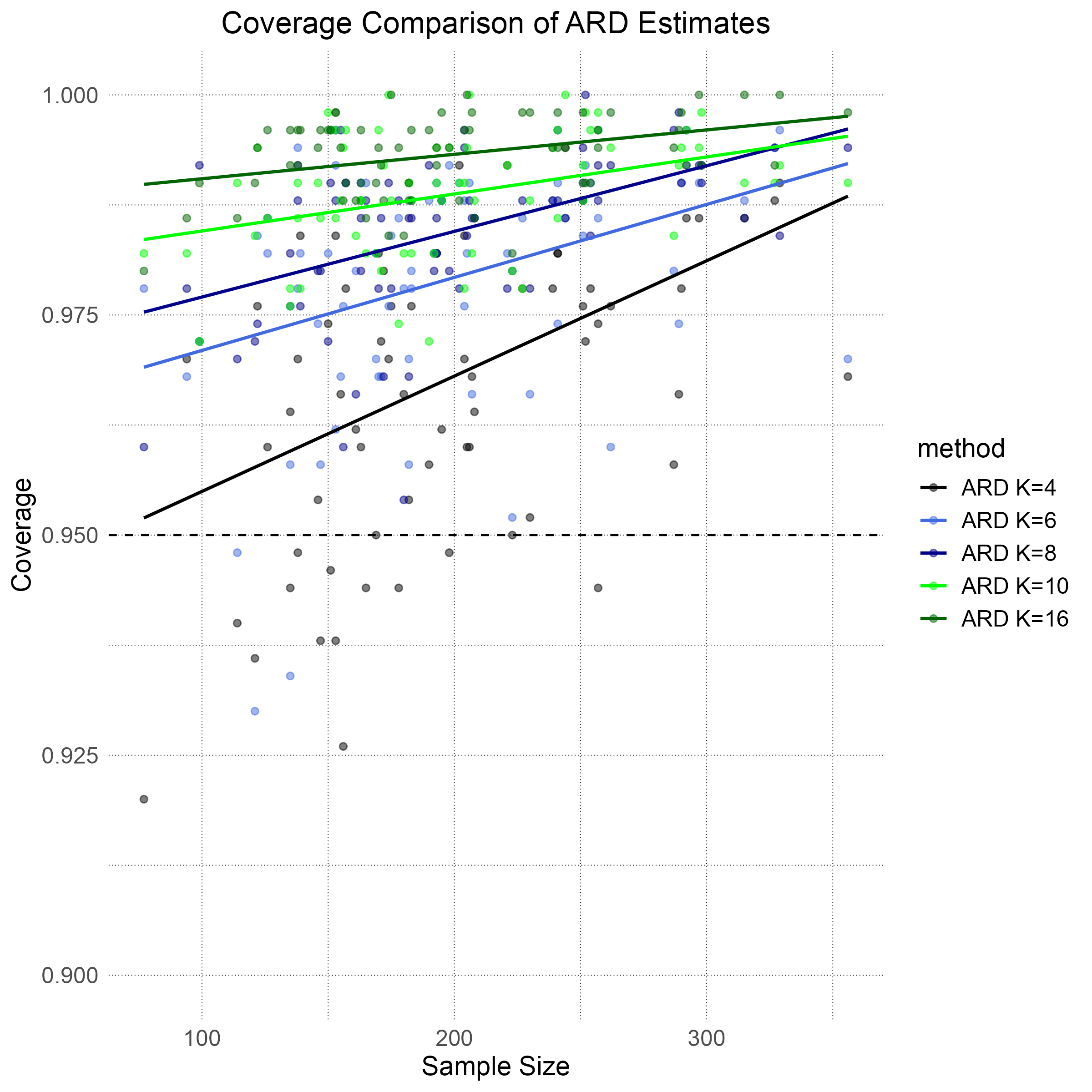}
    \caption{Coverage of the GATE using Eicker-Huber-White estimates of the variance. }
    \label{fig: Simulation GATE Coverage}
\end{figure}

In our simulation setup in section~\ref{sec: causal effect estimation} we can also compute confidence intervals based on the regression $Y_i = \beta^T \E[\tilde h(S_i, V_i)] + \epsilon_i$ where we apply the Eicker-Huber-White sandwich estimator of the variance. We then compute the corresponding plug-in estimator of the variance using the covariates observed and Lemma~\ref{lemma: plug in inference}. Since the covariates in the true regression model behave like averages over the graph, we expect Lemma~\ref{lemma: average parameter graph sample} to hold and therefore the difference between the GATE for any one draw of the graph, and the true GATE is very small. We see in Figure~\ref{fig: Simulation GATE Coverage} that the coverage tends to be larger than the nominal 95\%, though in general, due to model misspecification of the true-graph, there can be additional uncertainty due to the misspecification of the graph model. However, we see in this simple example that the coverage performs well with an off-the-shelf implementation.

\subsection{Experimental Design: Local Diffusion} \label{sec: example Local Diffusion}

\begin{figure}[htbp]
\centering
\subfigure[Bias of Full and Partial Data Diffusion Parameter Estimates]{
\includegraphics[width = 0.45\textwidth]{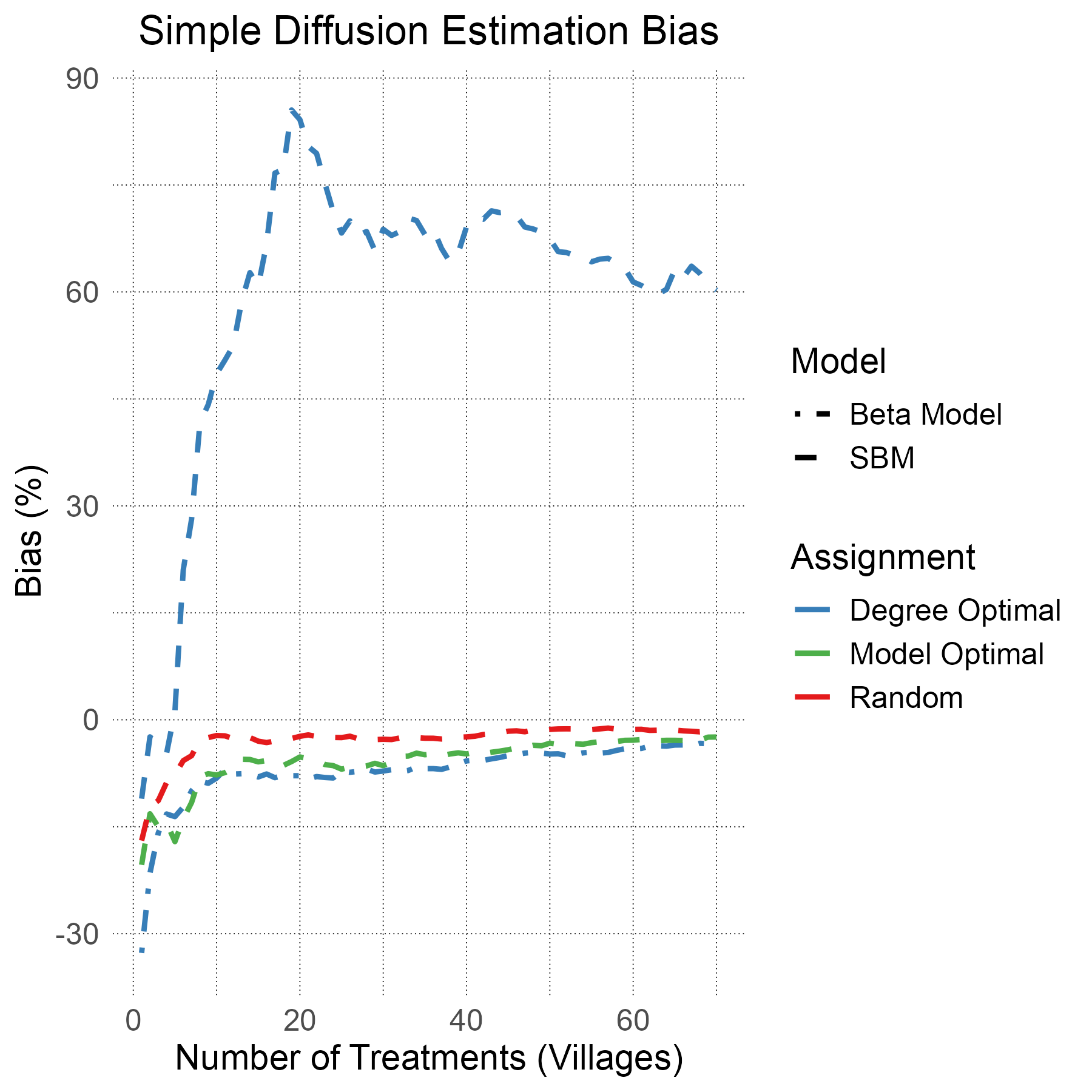}
}
\subfigure[RMSE of Full and Partial Data Diffusion Parameter Estimates]{
\includegraphics[width = 0.45\textwidth]{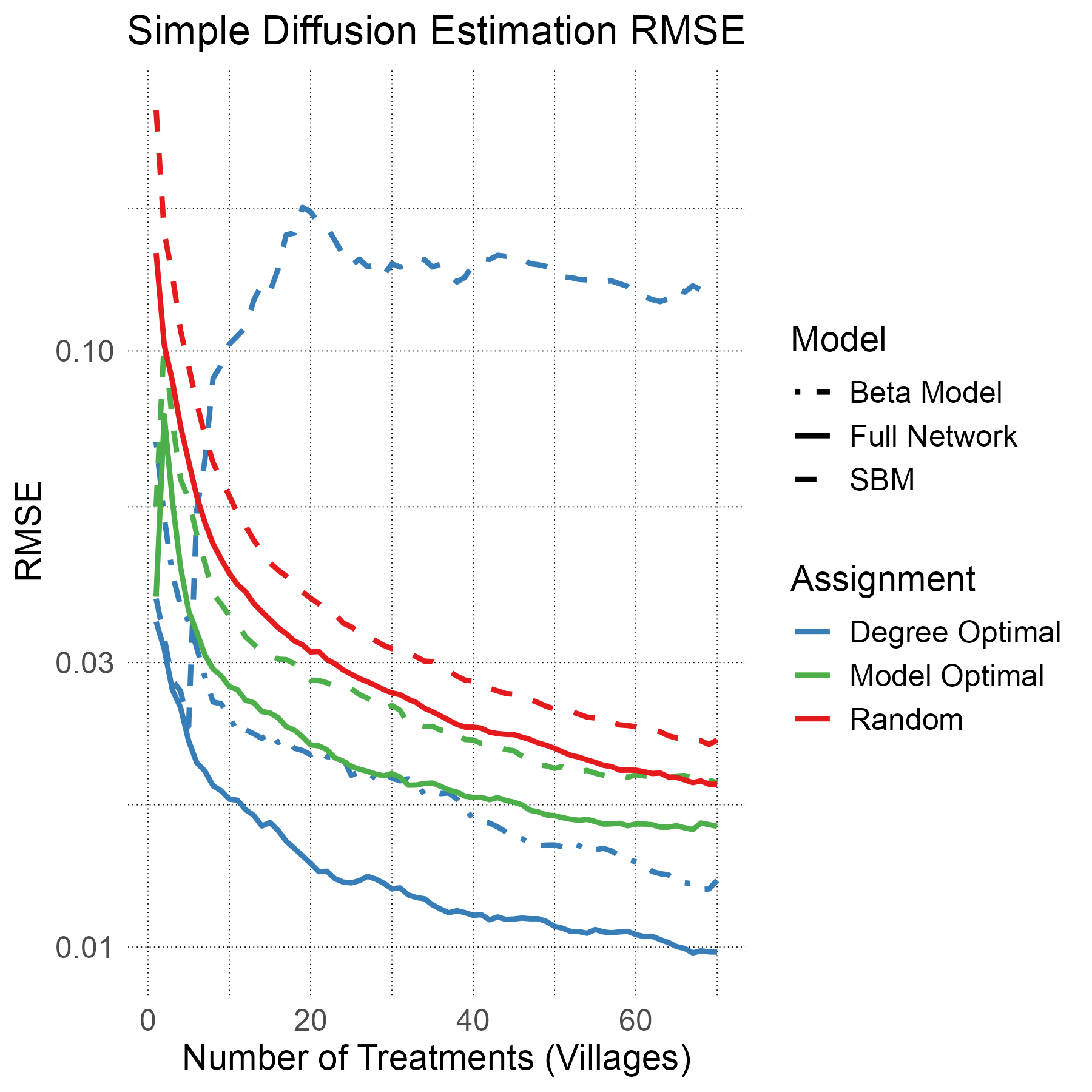}
}
\caption{RMSE and bias of estimating parameter $q$ using random seeding, and the optimal seed for each village. }
\label{fig: high local diffusion example}
\end{figure}

We next consider an example using a local diffusion process. We suppose that seed nodes are placed at time $0$ and that outcomes are measured at time $T = 1$, allowing for diffusion to only take place to the immediate neighbors with a fixed probability $q$. In this case, for non-seed nodes the probability of infection is related to the total number of treated neighbors through the following link function. Under this model let $V_i \in \{0,1\}$ denote the exposure as to whether one of their neighbors have received the treatment, i.e. $V_i = I(\sum_{j}G_{ij}a_j > 0)$. Then 
\al{
    \E[Y_i|V_i, S_i] &= q V_i 
}
In this experiment, a single individual is seeded in each network. Our goal is to identify the best individuals in each of the network to seed and rank them by the expected variance of the estimator. We compare this to random seeding of individuals in the network as well as seeding by only the highest degree nodes. We use the networks constructed by the union of all connections of \citet{banerjee2019gossip}. We construct estimates of the stochastic blockmodel as the partial data example using $K = 3$ in each case. We construct the traits using ARD responses based on number of connections with the following traits outlined in the Appendix in section~\ref{sec: Gossips ARD Questions}. We also include an alternative where a beta-model \citep{chatterjeed2011} is used in place of the SBM for the degree seeding where further details on estimation are included in section~\ref{sec: Estimation of the Beta Model}. We then draw samples of the graph using the parametric bootstrap to obtain a resampled distribution of ARD $\{\vect{X}^{*(b)}\}_{b= 1}^{B}$ for $B = 1000$. We identify the optimal treatment block for each parameter according to section~\ref{sec: Variance Minimization with Model Uncertainty}. We simulate $1000$ draws of the draws in the diffusion process for each true, and plot the associated bias and RMSE of the seeding strategies in Figure~\ref{fig: high local diffusion example} with a true diffusion parameter $q = 0.2$. 

In the full data case, the optimal strategy would be to seed the highest degree node in each of the networks and measure whether each of their neighbors are infected at time $T = 1$. However, this poses a problem for the stochastic blockmodel as we are essentially picking an outlier to seed, which is different than a typical member of the block over draws of the process. This can be corrected for using a model which accounts for degree heterogeneity, in our case, the beta model. In our optimal seeding strategy, we find that the RMSE is lower in both the degree optimized strategy with the beta model, as well as the block optimized strategy with the SBM, than even the full data version with a completely randomized allocation, hence highlighting the role of the interplay of the model of the graph and the experimental design. This behavior is observed in Figure~\ref{fig: high local diffusion example}.

\subsection{Estimated outcome model}  \label{sec: transportability for optimal seeding}
In this example, we consider a problem of optimal treatment assignment after the outcome model is estimated. We consider an example where an outcome model is estimated and transported to a new population. In this example we suppose that there is some benefit $\beta_1 > 0$ to receiving a treatment, and some smaller benefit based on the fraction of the neighbors treated $0 < \beta_2 < \beta_1$. We wish to assign treatments in a way that will maximize the expected outcome $\Psi(\vect{a}|G)$ for each network. 

\begin{equation*}
    Y_i = \beta_0 + \beta_{1} a_i + \beta_2 q_1 + \epsilon_i 
\end{equation*}

Where $q_i := \frac{1}{d_i}\sum_{j = 1}^n G_{ij}a_j$ denotes the normalized number of treated neighbors. We simulate the data with $\beta_0 = 1$, $\beta_1 = 1$ and $\beta_2 = 1/2$ with $\sigma_i \sim N(0,1)$.  We choose this form of a response function since it will be simple to solve with an off the shelf mixed-integer programming approach using CVXR \citep{Fu2020CVXR:Optimization}. 

We suppose that in each example there is only a budget for $B \in \{10,20,40,80\}$ treatments for each of the villages. The goal is to maximize the overall expected outcome. We consider the following competing procedures. In this case, we suppose that we have a single pilot network where we can learn the model and the goal is to maximize the benefit on the remaining networks. We use the same gossip diffusion networks as in sections~\ref{sec: example Local Diffusion} and \ref{sec: example Experimental Design Simulations}. 
%\ref{sec: example Generalized Hearing Model}. 

We compare the following seeding strategies. 
\begin{enumerate}
    \item Random assignment to all individuals in the network 
    \item Equal assignment amongst clusters. 
    \item Assign treatments ordered by the highest degree of the nodes. 
    \item Maximize the total expected outcome by maximizing $\max_{\vect{a}; \norm{a}_1 \leq B}\Psi(\vect{a}; \hat \beta, \hat \theta)$
\end{enumerate}

Let $\E[Y_i|\vect{a}] = \beta_0 + \beta_1a_i + \beta_2(1 - a_i) \sum_{k' = 1}^K \hat P_{\hat k_i k'} n_{t, k}$ and let $n_{t, k} = \sum_{j : k_j = k} a_j$. Therefore, the objective function. 
\al{
    \Psi(\vect{a}|\beta, \theta) &= \beta_0 + \frac{1}{n}\beta_1 \vect{1}^T\vect{n}_t + \frac{1}{n}\beta_2 \zeta^T\vect{n}_t 
}
where $\zeta = \frac{1}{d_i}\sum_{i = 1}^n \mathbf{P}_{k_i, \cdot}$ and $\vect{n}_t = (n_{t,1}, n_{t,2}, \dots, n_{t, K})$. In general, given a conditional model, one may fine tune the optimization approach to the particular challenges of evaluating the optimal treatment allocation. We partition each network into $6$ blocks. 

We plot the expected average outcome under each of the treatment allocations for the remaining $68$ networks after learning a model from the first pilot network. We repeat this for the total number of treatments $B \in \{10,20,40,80\}$. 

\begin{figure}
    \centering
    \includegraphics[height = 0.4\textheight]{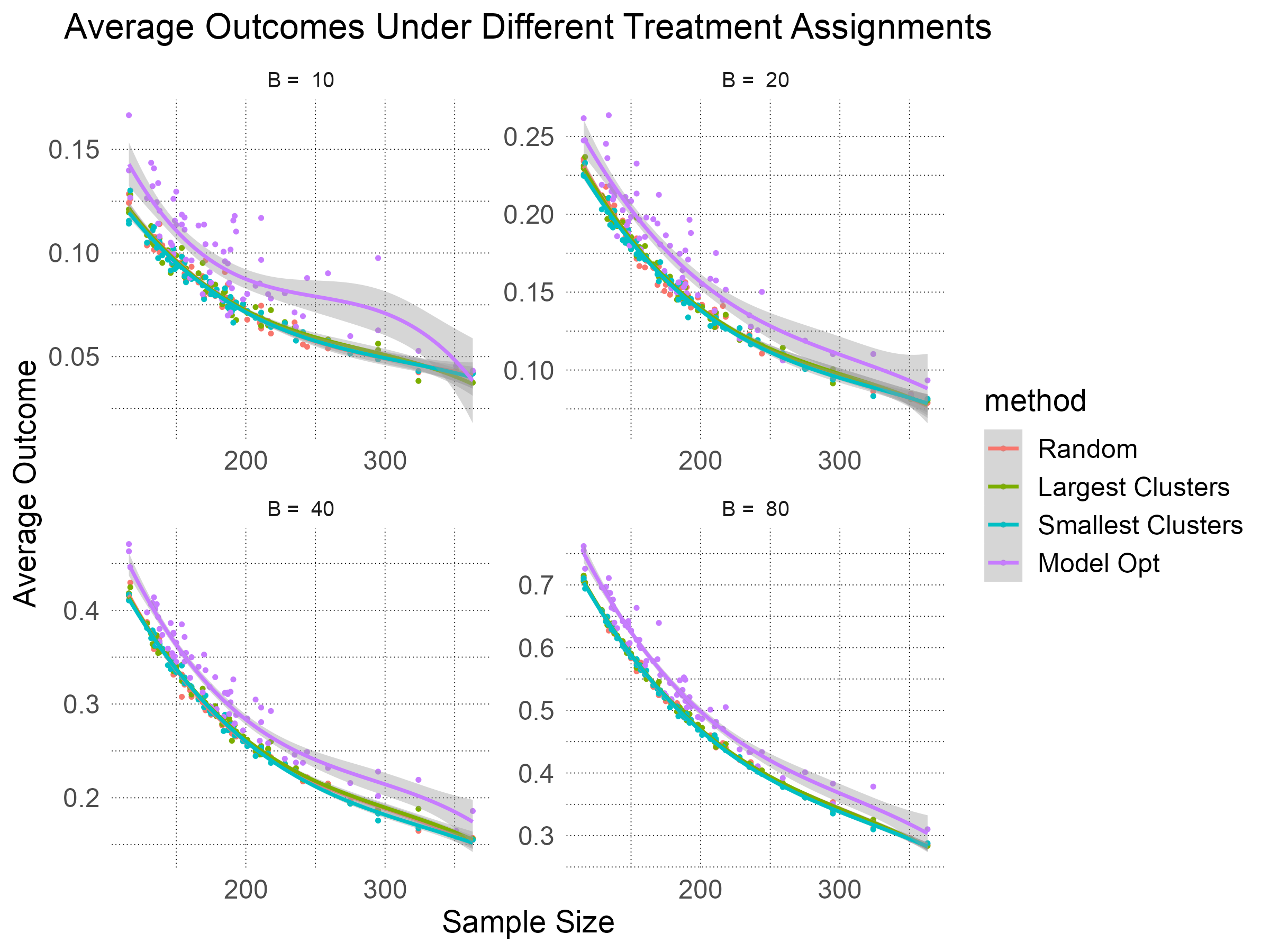}
    \caption{Our method, model based optimal treatment allocation (Model Opt) compared to random assignment and assignment to largest and smallest clusters respectively. The larger the values represent larger average outcomes in each of the networks. Curves are fit using cubic splines. The model based optimal design tends to give a higher value at each of the sample sizes at each treatment budget. For example, at a sample size of 150 and a treatment budget of 10, our methods leads to a $30\%$ increase in the average outcome. }
    \label{fig:example_optimal_treatment}
\end{figure}

In Figure~\ref{fig:example_optimal_treatment} we find that based on our method, we can achieve higher average outcomes than simple models based on the block positioning alone, emphasizing the importance of considering the potential outcome model when optimal targeting.

\subsection{Inference for evidence of complex contagion with partial network data} 
We can also replicate the results of \citet{beaman2021can}'s study on the evidence of pitplanting. They consider 3 measures of information diffusion. Firstly, if an individual has heard of pitplanting, second, if they know how to pitplant, and thirdly whether they adopt pitplanting in their practice. In order to control for one's position in the network, the authors consider the distance between the optimal seeds using two other targeting methods, simple diffusion, and geo-targeting as well as complex contagion. They then compare the increased odds of con 

\al{
    Y_{iv} &= \alpha + \beta_1 I(1 T Seeds) + \beta_2 I(2 T Seeds) + \beta_3 I(1 Simple)_{iv} + \beta_4 I(2 Simple)_{iv} \\
    & + \beta_5 I(1 Complex)_{iv} + \beta_6 I(2 Complex)_{iv} + \beta_7 I(1 Geo)_{iv} + \beta_8 I(2 Geo)_{iv} + \delta_{v} + \epsilon_{iv}
}

\begin{figure}
    \centering
    \includegraphics[width = 0.8\textwidth]{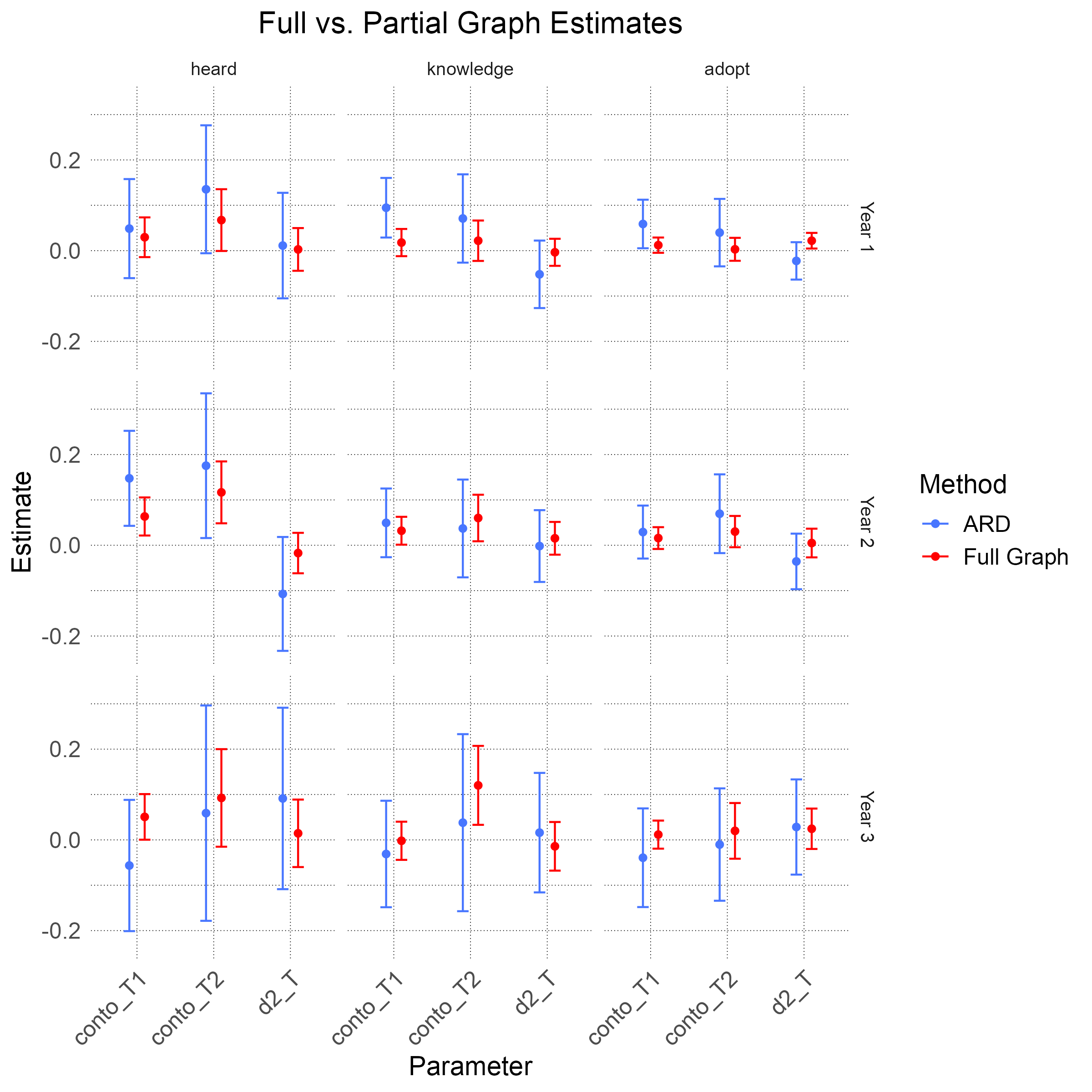}
    \caption{Replication of regression coefficients using aggregated relational data and associated 95\% confidence intervals. }
    \label{fig: Beaman Regression Tables 5 and 6}
\end{figure}

Again, we generate synthetic covariates and apply a stochastic blockmodel in order to estimate $K = 8$ blocks within each of the networks. We plot the coefficients for the connection to exactly $1$ seed, $2$ seeds and within radius $2$ of at least $1$ seed in Figure~\ref{fig: Beaman Regression Tables 5 and 6}. We note that we run the same regression as in \citet{beaman2021can}, however, some since the full network data includes some additional noise top preserve anonymity, we do not have the exact same estimates of the coefficients as in their paper, however, the conclusions are substantively the same. 

\subsection{Additional Experimental Details} \label{sec: Experimental Details}
To aid in reputability, we include additional details regarding the implementation of our methods as well as competing methods. 

\subsubsection{Beta Model Estimation} \label{sec: Estimation of the Beta Model}
Another common model utilized for random graph formation is the beta model coined by \citet{chatterjee2011random}. Namely these are a class of models that can be learned based on their degree sequence. We consider a version where each node has an affinity parameter $\nu_i$ and the probability of connection between each pair of nodes is $P(G_{ij} = 1) = \nu_i\nu_j$. Let $\nu_n = \sum_{i = 1}^n \nu_i$ Therefore, $\E[d_i = d] = \sum_{j \not = i} P(G_{ij} = 1) = \nu_i(\nu_n - \nu_i)$. The set of parameters $\{\nu_i\}_{i = 1}^n$ can be estimated using an iterative solution to the fixed point equation:
$$ \nu_i^{(t + 1)} = d_i/(\nu^{(t)}_n - \nu^{(t)}_i)$$

\subsubsection{ARD Questions} \label{sec: Gossips ARD Questions}
We utilize the measured traits to construct responses for ARD questions for each individual for the networks in \citet{banerjee2019gossip}. The constructed ARD include traits which ask "How many people do you know ..."
\begin{itemize}
    \item that are in each sub-caste? 
    \item that are Farmers, Shop owners, Domestic workers etc. ? 
    \item that own their house? 
    \item that have a house with at least 3 rooms? 
    \item that have access to electricity? 
\end{itemize}

For the estimation of the GATE using \citet{banerjee2013diffusion}, we use Leiden clustering and denote the clusters traits. When replicating the results of \citet{beaman2021can}, only a subset of nodes have available covariate. As was done in our examples with \citet{banerjee2013diffusion}, we construct synthetic traits using the clusters observed from Leiden clustering for $K = 10$. ARD is then constructed based on the connections to nodes of each trait.

\subsubsection{GATE Estimators}

The two estimators we compare for estimation of the global average treatment effect are the difference in means estimator $\widehat{\tau}_{DM}$ and the Horvitz-Thompson estimator $\widehat{\tau}_{HT}$. Let $E_{i0}$ and $E_{i1}$ denote the events that all neighbours of $i$ are untreated (including $i$ themselves) and treated respectively. 

\al{
    \widehat{\tau}_{DM} &= \frac{1}{n_1} \sum_{i = 1}^n Y_ia_i - \frac{1}{n_0} \sum_{i = 1}^n Y_i(1 - a_i) \\
    \widehat{\tau}_{HT} &= \frac{1}{n} \sum_{i = 1} \frac{Y_iI(E_{i1})}{P(E_{i1})} - \frac{Y_iI(E_{i0})}{P(E_{i0})}
}

In general, the Horvitz-Thompson estimator will be unbiased, however, it can often suffer from high variance for two reasons. Firstly, the probabilities of the events that all nodes are treated may be exceedingly low, inflating this variance, and also, relatively few nodes receive the exposures under which all of their neighbours are treated or none of them are.

In the case where the spillover effects are relatively mild, often a difference in means approach to the estimator is preferred. The effect of cluster randomization on the MSE of this estimator has been further studied in the complete network \citet{brennan2022cluster, viviano2020experimental}.

% \bibliographystyle{chicago} % Make sure that this is made to be lower case. 

% \bibliography{JASA_template_2021/references}
\end{document}